\documentclass[sigconf, nonacm]{acmart}

\usepackage{pvldb}
\usepackage{booktabs}
\usepackage{flafter}
\usepackage{tabularx}
\usepackage{fvextra}
\usepackage{tcolorbox}
\definecolor{promptink}{HTML}{1D2330}
\definecolor{promptframe}{HTML}{A7ABB2}
\definecolor{promptmuted}{HTML}{626B78}
\definecolor{promptarm}{HTML}{163F80}
\definecolor{promptarmbg}{HTML}{E6EEFA}
\newtcolorbox{promptbox}[1]{colback=white, colframe=promptframe, colbacktitle=promptink, coltitle=white,
  fonttitle=\sffamily\bfseries\scriptsize, title=#1, boxrule=0.5pt, arc=1.5pt, left=3pt, right=3pt,
  top=2pt, bottom=2pt, toptitle=1pt, bottomtitle=1pt, before skip=4pt, after skip=4pt}
\newtcolorbox{promptadd}{colback=promptarmbg, colframe=promptarmbg, boxrule=0pt, arc=1pt, left=2pt,
  right=2pt, top=1pt, bottom=1pt, before skip=2pt, after skip=2pt}
\usepackage{tikz}
\usetikzlibrary{arrows.meta}

\begin{document}
\title{Mind the Trap As LLMs Guess the Data Model: Grain-Aware Guards Against Fan-out in Text-to-SQL and Its Benchmarks
\unskip{}}

\author{Ahmed M. Mahran}
\orcid{0009-0007-8309-9331}
\affiliation{%
  \institution{Incorta}
  \country{Egypt}
}
\email{ahmahran@gmail.com}
\email{ahmed.mahran@incorta.com}

\begin{abstract}
Text-to-SQL systems built on large language models (LLMs) often aggregate over one-to-many and
many-to-many joins, a classic trap: when the joined rows repeat an entity, a SUM or COUNT counts it more
than once, yet the query runs and returns a plausible number. We characterize these \emph{fan-out errors} by one
condition on the relation an aggregate reads, which depends on two facts a schema does not show: each
measure's grain, and how many rows a join on each key repeats. An audit that tests the condition on
the data finds distorted gold answers for 4.2\% of the dev questions of BIRD, a widely used
benchmark, and BIRD's later, independent revision changed 44\% of the answers the audit flags as
definite fan-out, against 16\% of unflagged ones. On seven other benchmarks, distorted gold is no more common than in BIRD on public data, but affects 9--24\% of golds in two enterprise databases. We release BIRD-FanOut, 486 BIRD questions with reviewed gold, and GrainTraps,
synthetic questions that set this trap and related ones. Across eight LLM agents, functional dependencies (FDs) computed
from the data help no more than per-column statistics on GrainTraps, but grain-aware FDs, which also state each
measure's grain and the repetition per key, cut fan-out errors on BIRD from 18.4\% of runs to about
13\%, also on held-out questions and without LLM judges. Accuracy rises less, because most remaining
errors are not fan-out and BIRD's hints often prescribe the over-count. Text-to-SQL models should be given the grain of the data, and benchmark gold should be audited on
the data.
\par\smallskip\noindent Code, data and benchmarks are available at \url{https://github.com/ahmed-mahran/misqal}
\end{abstract}

\maketitle
\hypersetup{pdftitle={Mind the Trap As LLMs Guess the Data Model: Grain-Aware Guards Against Fan-out in Text-to-SQL and Its Benchmarks}, pdfauthor={Ahmed M. Mahran}}

\pagestyle{plain}

\section{Introduction}

\begin{figure}[t]
\centering
\includegraphics[width=\columnwidth]{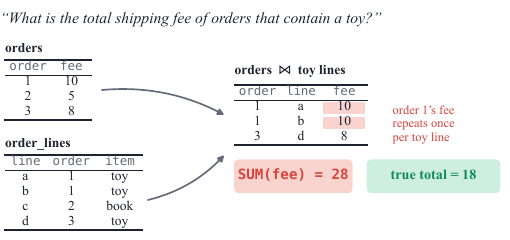}
\caption{A fan trap. The question needs the order lines to pick the orders, but the fee is stored once per order; the join repeats it once per toy line, so the sum counts order 1's fee twice (28 instead of 10 + 8 = 18).}
\label{fig:teaser}
\end{figure}

An analyst asks a text-to-SQL assistant for ``the total shipping fee of orders that contain a toy''.
The natural query joins the orders to their lines, keeps the toy lines and sums the fee (Figure~\ref{fig:teaser}). It runs and returns a plausible number, but it is wrong. The fee is stored once per order, the join
repeats it once per toy line, and an order with two toy lines is charged twice. The error
comes from an assumption about the data that the query never checked, namely that each order appears
once in the rows being summed.

This is one instance of a broad class of errors we call \emph{fan-out errors}. When a query aggregates
rows in which an entity repeats, a duplicate-sensitive aggregate (COUNT, SUM, AVG, or ORDER BY \ldots{}
LIMIT) sees that entity several times. Fan-out errors are the main kind of \emph{grain error}, an aggregate
computed at a grain other than the question's; the other kind groups or ranks at the wrong grain
without any repetition (\S{}2.1). The repetition has several sources: joins on non-unique keys
(the classic fan and chasm traps), denormalized tables that repeat order-level values on every order
line, and many-to-many mappings. Data warehousing guards against them by declaring the \emph{grain} of
each fact table, that is, what one row stands for (an order, an order line) \cite{Kimball}; semantic layers
declare join cardinalities \cite{Looker,MetricFlow,CortexAnalyst}, but these cannot describe dependencies
inside a single wide table. A text-to-SQL model
over a raw database has neither; it sees table and column names and must guess, or check, the data
model behind them.

\textbf{Fan-out is hard to see, for models and for evaluation.} A fan-out error raises no exception and
usually returns a plausible number. Evaluation rarely catches it. BIRD \cite{BIRD}, a widely used
text-to-SQL benchmark of 12,751 questions over 95 real databases, scores answers by set-based
execution accuracy (EX), which ignores duplicate result rows and so hides 5.8--6.8 points of
multiplicity errors on its dev set \cite{ModularSQL}. An over-count that changes an aggregate's value does
fail EX, but only against a correct gold, and BIRD's gold answers themselves contain fan-out; 4.2\% of
dev questions are distorted, and the cleaned release still has definite fan-out flags (\S{}4). Systems
that give the model cardinality information report small effects (removing it costs MCI-SQL 0.5 EX points \cite{MCI-SQL} and SQLfuse 2 points \cite{SQLfuse}), but these are measured by EX, not by fan-out errors.

\begin{figure*}[t]
\centering
\includegraphics[width=\textwidth]{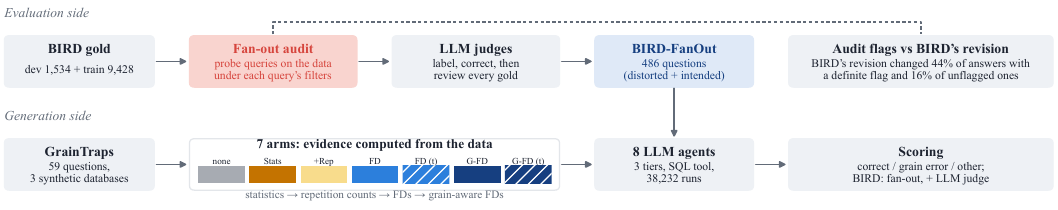}
\caption{The study. Top: an audit on the data flags BIRD gold queries that repeat the rows they aggregate; LLM judges label and correct the flags, BIRD's independent revision validates them, and the result is BIRD-FanOut. Bottom: eight LLM agents in three tiers answer BIRD-FanOut and GrainTraps, a separate synthetic trap suite, under up to seven evidence arms (six on BIRD dev; G-FD: \textsc{Grain-FD}, (t): delivered by a tool). These colours are used in every figure; hatched or hollow marks denote delivery through a tool. We count grain errors on GrainTraps (the answer reproduces the trap query's result or the audit flags it) and fan-out errors on BIRD (exact, or judged by an LLM; \S6).}
\label{fig:overview}
\end{figure*}

\textbf{This paper.} A text-to-SQL model needs two facts that a schema does not show, each measure's grain
and how many rows a join on each key repeats, and both can be computed from the data. We ask three questions. How much fan-out is in BIRD's gold, and can an audit that
tests on the data find it? When do current LLMs make fan-out errors? And does evidence computed from
the data help them avoid these errors, in particular evidence that states each measure's grain and how
many rows a join on each key repeats? Each model runs as an agent that queries the database through a
SQL tool and then submits one final query. We compare seven \emph{arms}, conditions that differ only in the
evidence the model receives. Figure~\ref{fig:overview} maps the study.

\textbf{Contributions and findings.} The paper makes four contributions, each with its main finding:

\begin{enumerate}
\def\labelenumi{\arabic{enumi}.}
\item
  \textbf{A grain condition for aggregates} (\S{}2.2). A duplicate-sensitive aggregate is correct when each
  entity at its measure's grain appears once in the relation it reads, and its error is the measure
  weighted by each entity's surplus multiplicity. This is a query-level form of OLAP summarizability
  \cite{LenzShoshani1997} that applies to the relation any SQL aggregate reads and can be checked on the
  data. It names the two facts a model must know, the measure's grain and the multiplicity the joins
  induce, and reads the fan, chasm and many-to-many traps as normal-form violations.
\item
  \textbf{A fan-out audit and BIRD-FanOut} (\S{}4). The audit tests on the data, under each query's own
  filters, whether the rows an aggregate reads repeat. Fan-out distorts 4.2\% of BIRD's dev gold, 6.2\% of train golds have a definite flag, and BIRD's independent revision changed 44\% of the gold answers the audit
  flags as definite fan-out, against 16\% of unflagged ones. On seven other benchmarks, distorted gold is
  no more common than in BIRD on public data but affects 9--24\% of golds in two enterprise databases.
  BIRD-FanOut has 486 BIRD questions whose gold was labelled,
  corrected and reviewed by LLM judges. Run on model answers, the audit flags most over-counts and few
  correct answers, but cannot tell whether a question meant to count the repeats (\S{}6.7).
\item
  \textbf{GrainTraps} (\S{}5.1), a synthetic suite separate from BIRD-FanOut: 59 questions on three generated
  databases that isolate six traps, in explicit and vague wordings. Without evidence, every model makes grain
  errors on it, in 2\% to 42\% of runs. The models that check the data before answering make the fewest, but
  telling a model to check makes it check more without reducing its errors (\S{}6.1).
\item
  \textbf{A controlled study of data-derived evidence} (\S{}5.2, \S{}6): eight LLMs, seven arms and 38,232 agent
  runs. On GrainTraps, FD labels such as ``can have many'' do no better than per-column statistics. Grain-aware FDs,
  which state each measure's grain, how many rows a join on each key repeats and where a blank breaks
  a dependency, cut fan-out errors on BIRD train from 18.4\% of runs without evidence, and 17.0\% with
  statistics, to 13.1--13.6\%, and the cut holds on 183 held-out questions; the repetition
  counts alone give about half of this cut. Accuracy rises only about 3 points, because BIRD's hints
  often prescribe the distorted formula and small over-counts barely show in the data (\S{}6.4).
\end{enumerate}

\section{Background}

\subsection{Grain errors and their traps}

\begin{figure*}[t]
\centering
\includegraphics[width=\textwidth]{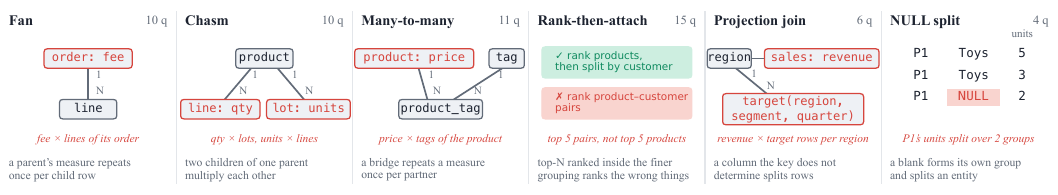}
\caption{Six traps that cause grain errors; all but rank-then-attach and NULL split are fan-out. Red marks the measure that repeats or the entity that splits, and the italic line says by how much; for rank-then-attach, the green box is the correct order of operations and the red one the trap. The number at the top right of each panel is the number of GrainTraps questions (\S5.1) of that type, explicit and vague wordings together; controls not shown.}
\label{fig:traps}
\end{figure*}

Figure~\ref{fig:traps} sketches six ways a query can count an entity more than once or at the wrong grain.
Fan and chasm traps are the connection traps of entity-relationship modelling, an ambiguous or a
broken path between entities \cite{Howe2001,ConnollyBegg2015}; BI tools use the names for the over-counts that joins
along such paths produce \cite{CortexAnalyst}, and we extend the name to the other four.
A \emph{measure} is a numeric column that queries aggregate (a fee, a quantity); its \emph{grain} is the entity
it has one value per (an order for a shipping fee, a line for a quantity). A question with a trap
usually has a \emph{trap query}, the natural SQL that falls into it.

\begin{itemize}
\item
  \textbf{Fan.} A parent's measure repeats once per child row, whether a join brings in the children or a
  denormalized table already holds them: an order's shipping fee counts once per line (Figure~\ref{fig:teaser}).
\item
  \textbf{Chasm.} Two independent children of one parent are joined through it and multiply each other:
  each order line of a product repeats once per inventory lot, and each lot once per line.
\item
  \textbf{Many-to-many mapping.} A bridge repeats a measure once per partner: a product's price counts once
  per tag.
\item
  \textbf{Rank-then-attach.} A top-N ranked inside a finer grouping than the question's ranks the wrong
  things: asked for the top five products and their sales per customer, the query returns the five
  largest product--customer pairs.
\item
  \textbf{Projection join.} A column that the grouping key does not determine splits or repeats rows: a
  region's revenue repeats once per target row of the region.
\item
  \textbf{NULL split.} A blank forms its own group and splits an entity: a product whose category is
  missing on some rows appears in two groups.
\end{itemize}

An answer that falls into a trap makes a \emph{grain error}: it aggregates at a grain other than the
question's. Grain errors are of two kinds:

\begin{itemize}
\item
  \emph{fan-out errors}: rows repeat an entity, as in fan, chasm, many-to-many and projection-join traps
  and in denormalized tables (the fan trap is the simplest way a join fans rows out);
\item
  \emph{grouping errors}: the grain is wrong but no row repeats, as in rank-then-attach and NULL splits.
\end{itemize}

The trap query makes its question's grain error by construction, so an answer whose result equals
the trap query's makes that error; we call such an error \emph{exact}. An answer that also makes another
mistake returns a different result, so \S{}5.1 also counts grain errors that the audit finds, and \S{}6 those an LLM judge finds.

\subsection{When is an aggregate correct?}

Let a query scope aggregate a measure \(m\) over the relation \(Q = \sigma_\varphi(R_1 \bowtie \cdots
\bowtie R_k)\) under bag semantics (within each group, if the scope groups), where \(m\) is a column of
some \(R_i\) with grain \(G\): the functional dependency \(G \to m\) holds in \(R_i\), and the question asks
for one value of \(m\) per \(G\)-value. Because every FD of an input holds in a selection and in an
equi-join, \(G \to m\) also holds in \(Q\). Let \(E\) be the set of \(G\)-values in \(Q\), and let the
\emph{multiplicity} \(\mu(g)\) be the number of tuples of \(Q\) with \(G = g\). In words, a duplicate-sensitive
aggregate over \(Q\) weights each entity by its multiplicity:

\begin{proposition}[Grain condition]
$\mathrm{SUM}_Q(m) = \sum_{g \in E} \mu(g)\, m(g)$ over the $g$ with non-NULL $m(g)$, so it differs
from the sum over entities by $\sum_{g} (\mu(g) - 1)\, m(g)$, and $\mathrm{COUNT}_Q(m)$ differs from
the number of entities with a non-NULL $m$ by $\sum_{g:\, m(g) \neq \mathrm{NULL}} (\mu(g) - 1)$.
Hence SUM and COUNT are correct when $\mu(g) = 1$ for every such $g$; COUNT is correct only then, and
SUM of a non-negative measure only if $\mu(g) = 1$ wherever $m(g) > 0$. AVG is the $\mu$-weighted
mean of $m$ over $E$, and it is correct for every $m$ if and only if $\mu$ is constant on $E$. MIN,
MAX and COUNT(DISTINCT $m$) do not depend on $\mu$, and ORDER BY \ldots{} LIMIT over a distorted
aggregate ranks by the distorted values.
\end{proposition}

The multiplicity traps of \S{}2.1 are the ways \(\mu\) exceeds one. In a fan trap, \(\mu(g)\) is the number
of child rows of parent \(g\). In a chasm trap, two independent children \(S\) and \(T\) of parent \(p\) give a
measure of \(S\) the multiplicity \(|T_p|\) and a measure of the parent \(|S_p| \cdot |T_p|\) (a parent
with an empty child drops out). A bridge contributes one tuple per partner, a projection join one per
value of the attached column, and a denormalized table has \(\mu > 1\) with no join at all. Assuming no
base table holds duplicate rows, \(\mu(g) > 1\) means that \(G\) is not a superkey of \(Q\) although \(G \to
m\) holds in \(Q\), so \(Q\) violates BCNF: fan-out is a normalization failure of the relation the
aggregate reads, a transitive dependency (line \(\to\) order \(\to\) fee) in the fan trap and a product of
independent multivalued facts in the chasm trap. Rank-then-attach and NULL splits are errors in the
grain of the grouping rather than in multiplicity, and fall outside the proposition. The condition
concerns the relation an aggregate reads, not the answer: users ask for breakdowns, and 9 of the 32
explicit gold answers of GrainTraps (controls included) are not in BCNF. Of an answer we require
only one value per cell, which the prompt also asks for (\S{}5.2).

The proposition is a query-level counterpart of summarizability in OLAP \cite{LenzShoshani1997,Mazon2009}, whose conditions (disjointness, completeness, type compatibility) are stated over declared
dimension hierarchies. Here the condition is stated for the relation an arbitrary SQL aggregate reads,
after its joins and filters, and both of its inputs can be measured on the data: the grain \(G\) of each
measure, and the multiplicity \(\mu\) that the query's joins and filters induce. Neither is visible in a
schema. The audit (\S{}4) tests on the data whether \(\mu > 1\) at the grain of each table's rows, under
the query's own filters. The grain-aware FD evidence (\S{}5.2) gives the model per-table forms of both
inputs: the grain of every measure, and, for each identifier, how many rows each value appears on,
which is the factor by which a join on it multiplies the matching rows; the model must still combine
them across its joins (\S{}2.3). An FD label such as ``each \texttt{order\_id} can have many \texttt{line\_id}'' says only
that \(\mu\) may exceed one, not by how much or for which measure.

\subsection{Instance-level functional dependencies}

Our dependency evidence (\S{}5.2) is computed on the data, so we use instance-level functional
dependencies (FDs). For columns \emph{A}, \emph{B} of a table \emph{r}, \emph{r} satisfies \emph{A $\rightarrow$ B} iff \emph{d(A) = d(AB)}, where \emph{d} counts
distinct value combinations \cite{TANE,BellBrockhausen}. We count NULL as a value (null = null, the
semantics of SQL GROUP BY); NULL-aware FD theory distinguishes other notions \cite{KoehlerLink,BertiEquille}. Comparing \emph{d(A)}, \emph{d(B)} and \emph{d(AB)} labels each pair 1:1, N:1, 1:N or M:N. An FD that
holds on the table holds under every filter; a pair without an FD (1:N, M:N) may gain one under a
filter, so its warning (``can have many'') is conservative. Per-table FDs carry over to joins by transitivity (\emph{A $\rightarrow$ id} in one table and \emph{id $\rightarrow$ B} in
the other give \emph{A $\rightarrow$ B} in the join) \cite{Klug,Paulley}, but non-dependencies do not, and whether a join
repeats rows depends on the uniqueness of its key. The models receive only per-table facts and must
combine them; stating the consequences for the model's own joins is future work (\S{}7).

\section{Related work}

\textbf{Data-derived dependencies for LLMs.} MCI-SQL \cite{MCI-SQL} mines within-table pair FDs with TANE and
exposes them through a metadata tool (removing it costs 0.5 EX on BIRD Mini-Dev). SQLfuse \cite{SQLfuse}
uses data-derived one-to-many relations (2.0 EX on Spider dev \cite{Spider}). Shkapenyuk et al.~\cite{Shkapenyuk}
give single-column profiles; DexterSQL \cite{DexterSQL} measures join fan-out to disambiguate columns; Su
et al.~\cite{Su} report little gain from cardinality-annotated keys. Our FD labels (\textsc{FD},
\S{}5.2) are in the spirit of MCI-SQL's pair labels, not a reproduction. Unlike these works, we measure
fan-out errors directly rather than EX, and separate statistics, repetition counts and dependencies
in controlled arms.

\textbf{Agentic exploration.} Agents that query the database before answering \cite{ReFoRCE,SpiderAgent}
check the data implicitly; we measure when they do so for fan-out.

\textbf{Ontologies and knowledge graphs.} Ontology-based access to relational data \cite{Ontop,ATHENA} lets a
question be posed over entities rather than tables. Sequeda et al.~\cite{Sequeda2023} show that asking
GPT-4 over a hand-built ontology and knowledge graph of an insurance database raises accuracy from
16\% to 54\%, and an ontology-based query check with LLM repair raises it to 72\% \cite{AllemangSequeda2024};
NVIDIA's Auto Ontology builds such a layer automatically \cite{AutoOntology}. A functional property in an ontology (``an order has exactly
one shipping fee'') is a declared dependency, and an ontology that models orders as their own class
prevents the fan trap by construction. Our evidence is the instance-level, automatically derived
counterpart of that part of an ontology, and our audit plays the role of the ontology-based query
check, testing the query against the data rather than against declared axioms.

\textbf{Fan-out detection and repair.} Semantic layers and their verifiers take cardinality from
declarations: Cortex Analyst's fan and chasm checks \cite{CortexAnalyst}, Looker's symmetric aggregates over
declared primary keys \cite{LookerSymmetric}, GROUND's declared grain
\cite{GROUND}, dbt linters \cite{grainguard} (in 14 public dbt projects, the grain of 463 of 765 joins is unknown to the linter).
DevRev's LLM plan verifier checks where deduplication is placed \cite{DevRev}; ModularSQL flags results with
many duplicate rows and adds DISTINCT to the outer query \cite{ModularSQL}, so an over-counted aggregate,
a single plausible number, passes it; a rule-based double-counting resolver predates LLMs
\cite{US11416488}. Grain Theory formalizes grain at the type level \cite{Karayannidis}, and summarizability \cite{LenzShoshani1997,Mazon2009} states when aggregation along declared dimension hierarchies is correct; eager aggregation
\cite{YanLarson,ChaudhuriShim} is classic. All of these rely on declared cardinality or grain, or patch
the query without looking at the data; our audit needs no declarations and tests each query on the
data.

\textbf{FD discovery.} TANE \cite{TANE}, HyFD \cite{HyFD} and a survey \cite{Abedjan} cover discovery; Abedjan and Naumann's unique-column discovery \cite{HCA} covers the
count-based pruning we use; Kossmann et al.~validate dependencies driven by queries \cite{Kossmann}. We
claim no novelty in computing the evidence; our question is which form of it helps a model.

\textbf{Benchmark quality.} Jin et al.~\cite{Jin} find annotation errors in over half of BIRD Mini-Dev and
Spider 2.0-Snow samples, including joins on non-unique keys; sqlsure reports an 8$\times$ inflated BIRD
gold \cite{sqlsure}; a practitioner analysis attributes part of a system's BIRD losses to over-counting
gold \cite{Hochman}; Wretblad et al.~\cite{Wretblad2024} find noisy questions and gold in a sample of BIRD.
Test-suite evaluation \cite{ZhongTestSuite} compares a query with the gold on many databases, so it still
trusts the gold, and ambiguity benchmarks \cite{AmbiQT,AMBROSIA} pair a question with several valid
readings, as our intended questions do. We audit eight benchmarks' gold on the data and check the audit
against BIRD's revision.

\section{Fan-out in BIRD}

\begin{figure}[t]
\centering
\includegraphics[width=\columnwidth]{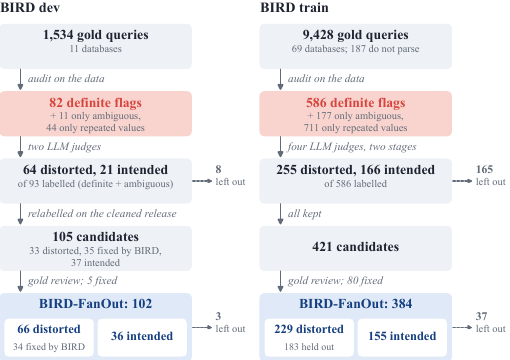}
\caption{From BIRD gold queries to BIRD-FanOut: 10,962 gold queries, 668 definite flags, 486 questions. Dev labels cover the definite and ambiguous flags of the June 2024 release; the dev candidates come from labelling BIRD's cleaned release. Left out: contested, benign or unresolvable questions and, in the gold review, disputed ones or ones without a fan-out signal. Fan-out errors are measured on the \textit{distorted} questions, whose original gold over-counts (183 of train's are held out from the design of \textsc{Grain-FD}); harm on the \textit{intended} ones, where counting the repeated rows is right.}
\label{fig:funnel}
\end{figure}

We audit every BIRD gold query on its data, check the audit against BIRD's own later revision, and
build BIRD-FanOut from the flags (Figure~\ref{fig:funnel}).

\begin{figure}[t]
\centering
\includegraphics[width=\columnwidth]{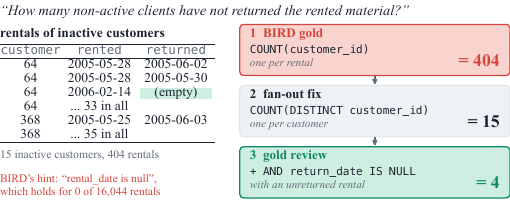}
\caption{BIRD train question 9279. (1) The gold counts rentals; the audit flags it as definite fan-out. (2) The fan-out correction counts customers. (3) The gold review adds the missing ``not returned'' condition. (empty) is NULL.}
\label{fig:goldfix}
\end{figure}

\textbf{The audit.} The audit tests whether the rows an aggregate reads repeat (\S{}2.2). It parses each
query (sqlglot \cite{sqlglot}, SQLite dialect) into scopes: CTEs, subqueries and branches of set
operations. For each scope with a duplicate-sensitive aggregate or an ORDER BY \ldots{} LIMIT, probe queries
under the scope's own filters compare the number of joined rows with the number of distinct rows of
each table (SQLite \texttt{rowid}). A join that could repeat rows but does not on this data is not flagged.
For example, the gold of BIRD train question 9279 (Figure~\ref{fig:goldfix}) counts \texttt{customer\_id} over
rentals joined to customers, so an inactive customer counts once per rental (404 rentals of 15
customers); the audit flags the COUNT as definite fan-out.

A flagged query gets
one of three flags:

\begin{itemize}
\item
  \emph{Definite}: rows repeat under SUM, AVG, COUNT(col) or ORDER BY \ldots{} LIMIT, or under COUNT(*) when every
  joined table repeats. The answer is wrong whatever the question means.
\item
  \emph{Ambiguous}: COUNT(*) where only some tables repeat. It counts the rows of the other tables exactly,
  so it is wrong only if the question counts a repeating one.
\item
  \emph{Repeated values}: COUNT(col) whose column repeats within its own table.
\end{itemize}

The audit takes about 16 s for BIRD dev. Within one denormalized table it checks only COUNT(col), a join to a
subquery or CTE is checked only statically, and AVG is flagged over any repetition, which is
conservative when rows repeat evenly (\S{}2.2).

\textbf{Prevalence.} On the June 2024 dev, 5.3\% of gold queries have a definite flag. Two judges (Claude
Opus 5.5 \cite{ClaudeOpus55}, GPT-5.5 \cite{GPT55}) label each definite or ambiguous flag as \emph{distorted} (the question means distinct entities), \emph{intended} or \emph{unclear}, write corrected SQL that must run and
pass the audit, and say whether the intended reading is \emph{resolvable} from the data model. They agree
on 95\% of flags, and a Claude agent reading 20 random flags agreed with each judge on 18. From the two labels, a flagged gold is

\begin{itemize}
\item
  \emph{distorted} if both judges label it so; its gold is corrected;
\item
  \emph{intended} if both label it intended: counting the repeated rows is right;
\item
  \emph{left out} if the judges disagree (\emph{contested}), the correction returns the same answer on this
  data (\emph{benign}), or the reading is not resolvable.
\end{itemize}

On dev, 4.2\% of golds are distorted; where the answer is one number, the distorted gold is a median
6.3 times the corrected answer. On train, 6.2\% of gold queries have a definite flag, and the distorted
ones are a median 1.6 times the corrected answer.

\begin{figure}[t]
\centering
\includegraphics[width=\columnwidth]{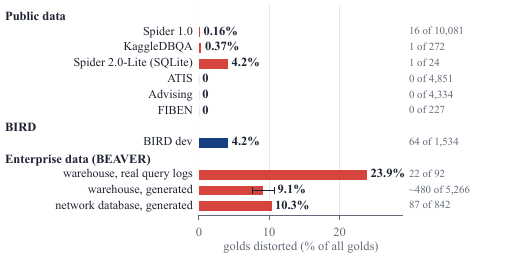}
\caption{Gold queries distorted by fan-out in eight benchmarks, as a share of all their golds (two Claude agents agree; BIRD: two LLM judges). Spider 1.0 pools dev, test and train; BEAVER's generated warehouse questions are scaled from 200 labelled flags (95\% interval).}
\label{fig:benchmarks}
\end{figure}

\textbf{Other benchmarks.} We ran the audit on the gold of seven other benchmarks that run on SQLite:
Spider 1.0 \cite{Spider}, KaggleDBQA \cite{KaggleDBQA}, the SQLite part of Spider 2.0-Lite \cite{SpiderAgent}, ATIS
and Advising \cite{FineganDollak}, FIBEN \cite{FIBEN} and the two databases of the enterprise benchmark BEAVER
\cite{BEAVER}, a data warehouse and a network database. Two Claude agents labelled each definite flag independently, running the gold and their
correction on the data, and agreed on 94\% of the labels; a gold is distorted if both find that the question
means each entity once and the correction changes the answer. Fan-out is rare in the academic
benchmarks even where most queries join, and no more common than in BIRD: at most 0.2\% of Spider's
golds are distorted, 0.4\% of KaggleDBQA's, 1 of the 24 SQLite golds of Spider 2.0-Lite, and none of
ATIS's, Advising's or FIBEN's. It is more common in BEAVER's two enterprise databases: 24\% of the
warehouse questions taken from real query logs (22 of 92), about 9.1\% of the warehouse questions
generated from templates, and 10.3\% of the network database's (Figure~\ref{fig:benchmarks}, Appendix D). Besides one-to-many
joins, duplicate keys in the warehouse's dimension tables repeat entities.

\begin{figure}[t]
\centering
\includegraphics[width=\columnwidth]{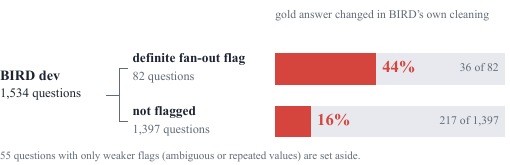}
\caption{The audit against BIRD's independent revision. Of the dev gold answers with definite fan-out flags, BIRD's cleaned release (2025), made without our flags, changed 44\%; of the unflagged ones, 16\%.}
\label{fig:cleaning}
\end{figure}

\textbf{Validation against BIRD's revision.} BIRD's cleaned dev (2025-11-06) \cite{BIRDDev2025} keeps the
question ids and rewrites gold queries and questions after human review, independently of our audit.
It changed the gold answer of 44\% of the questions with a definite flag, against 16\% of questions
with no flag (Figure~\ref{fig:cleaning}), and flagged questions were changed more often in each of the 9
databases that have flags. The release also shows how many real fan-out fixes the audit catches. Of the 36 answer changes that both judges call fan-out fixes keeping the question's meaning, the audit
flags 32. We looked for missed fixes only among changes that add DISTINCT, so the true recall may be
lower. The cleaned dev still has about 30 distorted golds by our labels; the 2026 Mini-Dev
\cite{BIRDMiniDev} has no definite flags.

\textbf{BIRD-FanOut.} BIRD-FanOut keeps the distorted and intended questions. Dev questions come from the
cleaned release; where the release fixed the fan-out, its SQL is the gold and
the old gold is the trap query. On train, two judges (Gemini 3.6 Flash \cite{Gemini36Flash},
Claude Haiku 5.5 \cite{ClaudeHaiku55}), chosen by agreement with the dev labels, label every definite
flag, and two strong judges (Claude Opus 5.5, GPT-5.6 Terra \cite{GPT56}) relabel those they leave
unsure. A distorted question is kept only if the two corrected queries return the same answer.

\textbf{Gold review.} A correction removes the fan-out but keeps the rest of BIRD's query, so a gold can stay
wrong for another reason: of 40 model answers that both judges called reasonable readings yet that did
not match the corrected gold, a Claude agent querying the data found the gold wrong in 21 (a negated
filter, a typo copied from the hint, a join on the wrong key). In question 9279 the gold also never checks that the rental was not returned (BIRD's hint puts this on
\texttt{rental\_date}, which is never empty); the review's fix lowers the answer from 15 to 4. Every gold of BIRD-FanOut is therefore
reviewed. Gemini
and Haiku read each question with its hint, the schema, sample rows and the gold's result;
anything either flags goes to the strong judges, who see each other's review, unnamed, once when they
disagree. A fix is accepted only if both strong judges call the gold wrong and their fixed queries
return the same result; it must run, change the gold's result and, on a distorted question, pass the
audit, and a fix that returns nothing, 0 or NULL is also checked on the data. Each judge also writes a
trap query, kept when both agree, so that exact fan-out errors can be scored without a judge. BIRD-FanOut
thus has 486 questions, 295 distorted and 191 intended. Appendix D gives the review's
outcome, its checks on the data and its effect on the results.

\section{GrainTraps and the evidence}

\subsection{GrainTraps}

BIRD shows how often fan-out occurs; GrainTraps shows when models make grain errors. It uses three
synthetic SQLite databases, generated reproducibly: \texttt{bookings} (one wide sales table; products booked under
several divisions and by several customers), \texttt{supply} (orders and inventory lots, both ``many'' per
product), and \texttt{shop} (a wide order-line table with order-level fields repeated per line, NULL gaps in
an attribute, many-to-many tags, and sales targets per region, customer segment and quarter). It has 59 questions: 29
explicit trap questions, 27 vague twins and 3 controls (questions per trap type in Figure~\ref{fig:traps}).

Explicit questions state the intended grain; vague twins drop it from the wording but keep the list
of returned columns, and each records why its reading follows from the data model. Every question
has a written trap explanation and solution, gold SQL, a trap query (the naive SQL that reproduces the trap),
and (for one question) an alternative accepted reading. The NULL-split gold treats the gaps as missing
values of a product's single category; reporting NULL as its own category is the trap reading.

\textbf{Scoring.} An answer is

\begin{itemize}
\item
  \emph{correct} if its rows match the gold or an alternative as multisets (floats to 2 decimals, order
  ignored);
\item
  a \emph{grain error} if it is exact or the audit (\S{}4) finds realized fan-out in it;
\item
  \emph{other} if it is wrong otherwise;
\item
  \emph{failed} if no runnable SQL was produced.
\end{itemize}

A model that ends without submitting is scored on the last query it ran; in the runs of \S{}6 this happens in at
most 4.0\% of runs of any model and arm.

\textbf{Role and construction.} We use the suite as a diagnostic rather than a leaderboard, since the four larger models of the full-schema runs (\S{}5.2) solve 92\% of
it without help. A Claude agent generated the databases, questions, gold and trap queries under the
authors' direction. Tests check on the data that every gold returns rows and every trap query returns
a different result; vague questions were kept only if strong models could solve them as written (in
runs without evidence), so strong models leave evidence little room to help. There is
no held-out split; the three controls are simple, and intended fan-out (counting joined rows on
purpose) is covered by BIRD-FanOut's intended questions instead.

\subsection{The evidence}

\begin{figure*}[t]
\centering
\includegraphics[width=\textwidth]{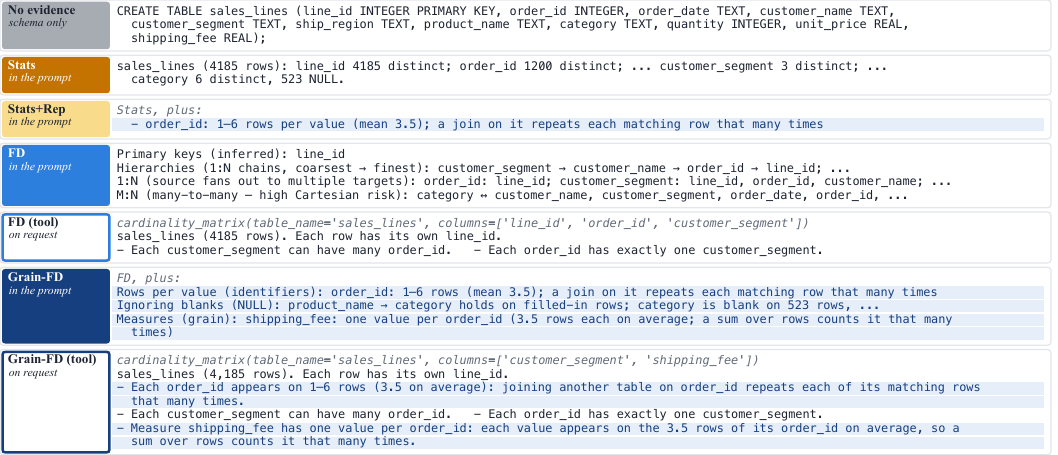}
\caption{What each arm gives the model for one GrainTraps question (``How much was charged in shipping fees in total for each customer segment?'', table \texttt{sales\_lines}), copied from the run traces and abridged (\dots); markdown marks are dropped and the many-to-many arrow is drawn as $\leftrightarrow$. Blue: what an arm adds over its reference arm; grey italics: the model's own tool call. Only the Grain-FD arms say that a shipping fee repeats on the 3.5 lines of its order.}
\label{fig:ladder}
\end{figure*}

\textbf{Agent and arms.} Each model runs as an agent: it can run SQL on the database through a tool as
often as it needs (up to 50 steps), and it ends by submitting one SQL query and a short answer
(Appendix B); we call the record of a run its \emph{trace}. All arms share one prompt, which says \emph{Do not
assume how tables and columns relate from their names or from what seems typical: check it in the
data}, and asks for exactly the requested columns with one value per cell. It contains no procedure,
trap names or layout hints. The seven arms differ only in the evidence the model receives, and
Figure~\ref{fig:ladder} shows, verbatim, what each arm gives the model for one GrainTraps question
(Appendix A shows a second).
\textsc{Stats} is the content control: rows, distinct values and NULLs per non-measure column, without relating columns.
\textsc{Stats+Rep} is an exploratory control that isolates the repetition counts. Evidence is given either in
the prompt or through a tool, \texttt{cardinality\_matrix}, that the model calls with a table and columns
and must call on every table its final query reads. The rest of this section describes the setting all arms share and how the evidence is computed.

\textbf{Known tables.} For every question, in every arm, the agent sees only the tables read by its
reference queries (the corrected gold and the trap query, or the original gold), with all their
columns, never the whole database. Keeping both queries' tables leaves both readings open. On BIRD
train this cuts the schema to 27\% of its size.

\textbf{Full-schema runs.} A separate set of runs shows the agent the whole database instead, with measure
columns included in the evidence: Claude Opus 5.5, Claude Sonnet 5.5, GPT-5.5, Gemini 3.1 Pro
\cite{Gemini31Pro}, Gemini 3.5 Flash-Lite, GLM, GPT-OSS-120B \cite{GPTOSS} (which did not complete the main runs) and Nemotron, on GrainTraps, BIRD-FanOut dev
and 150 ordinary Mini-Dev questions (2026 release, with joins and no audit flags). \S{}5.1, \S{}6.1, \S{}6.6 and \S{}6.7 use these runs;
Appendix D gives their arms and main results.

\textbf{Column roles.} Before any question is asked, every column of the known tables (2,879 columns in
352 tables) is labelled \emph{identifier}, \emph{dimension} or \emph{measure}, as in a data catalog. A heuristic
(keys, types, numeric share, BIRD's column descriptions) and two LLM judges per table (Gemini 3.6
Flash, Claude Haiku 5.5; they agree on 97\%) assign the roles, and Claude Sonnet 5.5 decides the
remaining columns; 704 columns are measures. The heuristic alone leaves 613 integer columns open and
misreads 116 numeric columns stored as text. The roles decide what the evidence covers. Measures are
left out of the statistics and the dependency profiles, because profiling them adds noise such as
accidental keys (``unit\_price is a key'' because all prices differ); \textsc{Grain-FD} instead
states each measure's grain. Identifiers get the repetition counts. The schema shows no roles; the
model learns that a column is a measure only from evidence that says so. The roles ship with the
benchmarks.

\textbf{FD evidence.} For a table and its non-measure columns, aggregate SQL computes \emph{d($\cdot$)} for
single columns and pairs, pruning pairs whose label follows from others (1:1 substitution,
transitivity, cardinality bounds, sampled counterexamples) \cite{HCA}. The profile reports the inferred
key, 1:N hierarchies and every pair's label (Figure~\ref{fig:ladder}).

\textbf{Grain-FD evidence.} The grain-aware evidence supplies the two inputs of \S{}2.2 with four additions
to the FD evidence. It is computed once per table, cached, and delivered exactly as the FD evidence,
so \textsc{FD} vs \textsc{Grain-FD} and \textsc{FD (tool)} vs \textsc{Grain-FD (tool)} differ only in content. The examples are
from the table of Figure~\ref{fig:ladder}.

\begin{itemize}
\item
  \emph{Measure grain.} Leaving measures out of the dependency profiles (above) also hides that a shipping fee
  repeats on every line of its order. \textsc{Grain-FD} states the grain of each measure: the coarsest
  non-measure column on which the measure has a single value (``\texttt{shipping\_fee}: one value per
  \texttt{order\_id} (3.5 rows each on average; a sum over rows counts it that many times)''), or, if no single column determines it, that it varies by row.
\item
  \emph{Repetition counts.} ``Can have many'' says that a join may repeat rows, not how often. \textsc{Grain-FD}
  adds, for each identifier, the number of rows per value and the consequence: ``\texttt{order\_id}: 1--6 rows
  (mean 3.5); a join on it repeats each matching row that many times''.
\item
  \emph{NULL-aware FDs.} FDs count NULL as a value (GROUP BY semantics), so ``each product has one
  category'' fails as soon as some rows leave the category blank, and the label reads ``can have many''.
  \textsc{Grain-FD} re-checks each failed FD on filled rows only and reports the gap: ``\texttt{product\_name} $\rightarrow$
  \texttt{category} holds on filled-in rows; \texttt{category} is blank on 523 rows''.
\item
  \emph{Scoping.} \textsc{Grain-FD (tool)} always adds the table's identifiers to the requested columns (up to
  25), so a scoped call cannot omit the key that carries a measure's grain.
\end{itemize}

\textbf{Design set and held-out questions.} The Grain-FD rules are generic, with no question-specific
content. We developed them on a pilot drawn from GrainTraps and from 46 BIRD-FanOut train questions
selected because \textsc{FD} evidence changed a model's outcome on them (Appendix D); all of GrainTraps and
these 46 questions form the \emph{design set}. The other 183 distorted train questions are \emph{held out}: they
played no part in the design. They are the complement of a set selected on \textsc{FD}'s outcomes, not a
random split.

\textbf{What each comparison measures.} \textsc{Stats} against no evidence gives what simple statistics add; each
FD arm against \textsc{Stats} gives what dependencies add beyond them, and each grain-aware arm against the FD
arm with the same delivery gives what grain-awareness adds. In the exploratory control,
\textsc{Stats+Rep} against \textsc{Stats} gives what the repetition counts add, and \textsc{Grain-FD} against \textsc{Stats+Rep}
roughly what the rest of the grain-aware evidence adds (roughly, because \textsc{Grain-FD} lacks the
per-column counts of \textsc{Stats+Rep}). \textsc{FD} and \textsc{FD (tool)} differ in both wording and delivery, so each is
compared with \textsc{Stats}, not with the other. The grain-aware arms are also the only ones that name the
measure columns, so their effect includes that labelling.

\section{Results}

\textbf{Models and runs.} We test eight models and group them by their grain errors on GrainTraps without
evidence: \emph{low-error} (Claude Sonnet 5.5 \cite{ClaudeSonnet55}, Claude Haiku 5.5 \cite{ClaudeHaiku55}; 2.4\%),
\emph{mid-error} (DeepSeek V4.1 Flash \cite{DeepSeekV41}, GPT-5.6 Luna and Terra \cite{GPT56}, GLM-5.3-Flash
\cite{GLM53}; 13.2\%) and \emph{high-error} (Gemini 3.5 Flash-Lite \cite{Gemini35FlashLite}, NVIDIA Nemotron 3.5 Super
VL 120B-A12B \cite{Nemotron35}; 40.8\%). BIRD-FanOut, which the grouping does not use, gives the same order
(fan-out errors without evidence 13.8\%, 17.7\%, 24.5\%; each step p $\le$ .004). The tiers only group the
results; on GrainTraps, a tier's change from no evidence is partly regression to the mean,
since the tiers were chosen on that arm. Each model answers every trap question three times and every BIRD-FanOut question once, in the six
main arms and, on GrainTraps and BIRD train only, in \textsc{Stats+Rep} (questions in
Figure~\ref{fig:funnel}). Runs were made on the BIRD questions as first labelled (421 train, 105 dev), except \textsc{Stats+Rep}, which ran
after the gold review on the 384 train questions it kept; the scores use only the 486 kept questions. There are 38,232 runs in total, one per question, arm and model on BIRD and three on GrainTraps, with
provider-default temperature and up to 50 agent steps. Three further models (Gemini 3.8 Flash, Qwen 3.8 27B, GPT-OSS-120B) did not complete every arm and
are excluded.

\textbf{Statistical tests.} Each evidence run is paired with its reference arm's run on the same question, model
and repeat. We report each comparison as the errors the tested arm avoids (grain errors on GrainTraps,
fan-out errors on BIRD), in percentage points of runs (positive is better). The unit of
inference is the question. On GrainTraps, a \emph{base question} is an explicit question together with its
vague twin; on BIRD, each question is its own base. Figures give 95\% bootstrap intervals over base
questions \cite{EfronTibshirani1993}; p-values are from a sign-flip permutation test on the per-question
changes unless marked as sign tests. The six headline tests (\textsc{Grain-FD} and \textsc{Grain-FD (tool)} against
\textsc{Stats}, on GrainTraps, on BIRD train and on its held-out questions) survive Holm correction (largest
adjusted p = .012). The eight tests involving \textsc{Stats+Rep} are exploratory; none survives Holm
correction within that family (smallest adjusted p = .058). Results by tier or by model are
descriptive.

\textbf{Measuring fan-out on BIRD.} All runs are scored against the \emph{reviewed golds} of \S{}4. Because the
audit that found BIRD-FanOut's questions detects repeated rows only, every distorted gold over-counts,
so on BIRD we measure fan-out errors rather than all grain errors. An answer that over-counts and also
makes another mistake is not an exact fan-out error. To catch such answers, an LLM judge that is not a
subject model labels every other distinct wrong answer as fan-out or not, and a second judge checks a
random sample and settles every disagreement with the audit (\S{}6.3). We report
three metrics of fan-out errors on BIRD:

\begin{itemize}
\item
  \emph{exact + judged} (primary, called \emph{fan-out errors}): exact fan-out errors plus answers judged to over-count;
\item
  \emph{exact only}: no LLM judge;
\item
  \emph{exact + audit flags}: no LLM judge; the rule used on GrainTraps (Figure~\ref{fig:robust}).
\end{itemize}

\subsection{Who makes grain errors?}

\begin{figure}[t]
\centering
\includegraphics[width=\columnwidth]{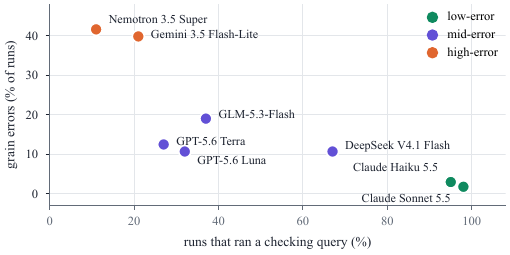}
\caption{GrainTraps without evidence: share of runs that ran a checking query and share with a grain error, per model (168 non-control runs each); colour gives the tier. Models that check more make fewer grain errors: the two Claude models check in 95--98\% of runs and almost never make one.}
\label{fig:check}
\end{figure}

Every model makes grain errors on GrainTraps, and the models that check the data make the fewest (Figure~\ref{fig:check}). Without evidence, grain errors range from 2\% (Sonnet) to 42\% (Nemotron) of the 168
non-control runs per model (Appendix C); the low-error models fail only on NULL splits and fan traps,
and solve every chasm question.

\emph{Checking.} A checking query is one the model ran before answering (other than its submission) that
tests how columns relate (COUNT DISTINCT, HAVING COUNT, IS NULL). The two Claude models check in 95--98\% of runs and almost never make a grain error; the others check
far less. Within a model, runs that checked make fewer grain errors (Terra 7\% vs 15\%, GLM 13\% vs 23\%,
Flash-Lite 26\% vs 44\%), but none of these differences is significant, either per model or when only
runs on the same question are compared (Mantel-Haenszel \cite{MantelHaenszel1959} p \textgreater{} .7). This suggests that checking marks the careful models rather than making a model careful. In the full-schema runs (\S{}5.2), adding the prompt's line
telling models to check the data to an earlier prompt raised checking sharply (GPT-5.5 3\% $\rightarrow$ 21\%, Sonnet 57\% $\rightarrow$ 94\%) without
reducing grain errors.

\subsection{Evidence on GrainTraps}

\begin{figure*}[t]
\centering
\includegraphics[width=\textwidth]{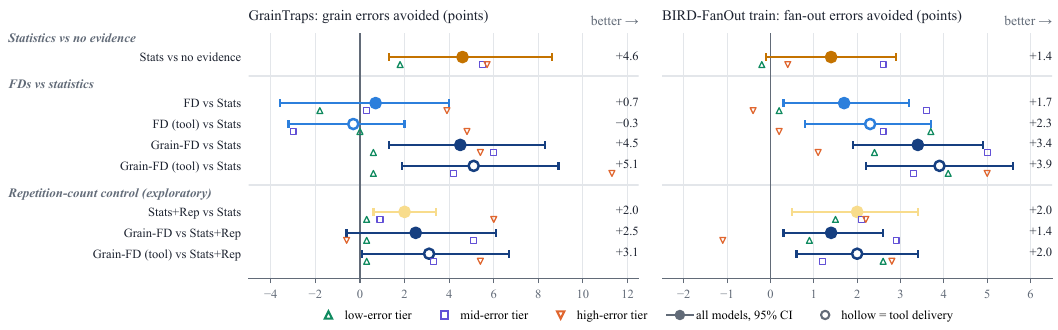}
\caption{Errors avoided by the tested arm relative to its reference, in percentage points of runs (right is better): grain errors on GrainTraps (in-sample for the Grain-FD arms) and fan-out errors on BIRD-FanOut train (exact plus judged; 229 questions). Large dots: all eight models with a 95\% bootstrap interval over questions (a trap question and its vague twin count as one); small markers: tiers (descriptive); hollow: tool delivery. On GrainTraps \textsc{FD} adds nothing over the statistics and on BIRD 1.7--2.3 points; the Grain-FD arms add most. Full tables, with accuracy, are in Appendix E.}
\label{fig:effects}
\end{figure*}

\textbf{On GrainTraps, statistics help, FDs add nothing, and grain-aware FDs avoid about five more points of
grain errors than the statistics} (Figure~\ref{fig:effects}, left). GrainTraps belongs to the design set,
so these results show that the evidence fixes the failures it targets; the held-out test is on BIRD
(\S{}6.3).

Figure~\ref{fig:effects} shows the paired differences between arms; each arm's absolute rates are listed
below (Table~\ref{tab:1}).

\begin{table}[t]
\centering\footnotesize
\caption{Error rates per arm, in \% of runs: grain errors on GrainTraps (1,344 runs per arm), fan-out errors on the 229 distorted BIRD-FanOut train questions (1,832) and on the 183 held-out ones (1,464), and accuracy on the 229. Bold: the best value in each column; \textsc{Stats+Rep} is exploratory.}
\label{tab:1}
\begin{tabular}{@{}lrrrr@{}}
\toprule
\textbf{arm} & \textbf{GrainTraps} & \textbf{BIRD train} & \textbf{held out} & \textbf{BIRD acc.} \\
\midrule
no evidence & 17.4 & 18.4 & 16.1 & 60.3 \\
\textsc{Stats} & 12.8 & 17.0 & 15.1 & 61.2 \\
\textsc{Stats+Rep} & 10.8 & 15.1 & 13.9 & 62.3 \\
\textsc{FD} & 12.1 & 15.3 & 13.9 & 62.0 \\
\textsc{FD (tool)} & 13.1 & 14.7 & 12.9 & 62.0 \\
\textsc{Grain-FD} & 8.3 & 13.6 & 12.4 & \textbf{63.3} \\
\textsc{Grain-FD (tool)} & \textbf{7.7} & \textbf{13.1} & \textbf{12.2} & 63.2 \\
\bottomrule
\end{tabular}
\end{table}

Grain errors fall from 17.4\% of runs without evidence to 12.8\% with the statistics (p = .015).
\textsc{FD} and \textsc{FD (tool)} add nothing beyond them; the low-error models even lose a little, all on NULL
splits, where an FD counts a blank as a value. \textsc{Grain-FD} and \textsc{Grain-FD (tool)} bring the rate down to
8.3\% and 7.7\%, 4.5 and 5.1 points below the statistics (p = .012 and .003), and accuracy rises by about
as much. Most of the gain is on the fan traps, the failure
the measure's grain targets: 5 of the 29 base questions give 35 of the 60 avoided grain errors.
Because the gain is concentrated, a sign test, which counts each base question once whatever its
change, is weaker (p = .078 and .041). \textsc{Grain-FD} leaves four of the six low- and mid-error models with no
grain error; the high-error models keep 39 and 61 grain errors in their
168 runs (Appendix C).

\emph{Repetition counts or FDs? (exploratory)} The repetition counts alone (\textsc{Stats+Rep}) avoid 2.0 points
of grain errors beyond the statistics (p = .019), 45\% of \textsc{Grain-FD}'s cut, but barely move accuracy.
\textsc{Grain-FD} avoids a further 2.5 points (not significant), mostly on the fan traps, where only it states
the measure's grain (21 grain errors against 49); on chasm questions it has more
(20 against 10).

\subsection{Real data: BIRD-FanOut}

\textbf{On BIRD train, grain-aware FDs cut fan-out errors from 18.4\% of runs to 13.1--13.6\%, 3.4--3.9 points
fewer than with the statistics (p \textless{} .001), also on held-out questions and when errors are counted
without any LLM judge.}

Without evidence, 12.9\% of runs on the 229 distorted train questions return exactly the distorted
gold's result. Counting also the answers that over-count but make another mistake as well, 18.4\% of
runs are fan-out errors (Table~\ref{tab:1}). Against the statistics,
\textsc{Grain-FD} avoids 3.4 points and \textsc{Grain-FD (tool)} 3.9 (both p \textless{} .001; Figure~\ref{fig:effects}, right), plain
FDs 1.7 and 2.3 points, and the statistics alone avoid 1.4 points against no evidence (not
significant). Against plain FDs with the same delivery, the grain-aware additions avoid a further 1.6
points in both deliveries ({[}0.3, 3.1{]} and {[}0.2, 3.1{]}; p = .03 and .04, outside the six headline tests);
on held-out questions, 1.6 (p = .05) and 0.7 points (not significant).

\begin{figure}[t]
\centering
\includegraphics[width=\columnwidth]{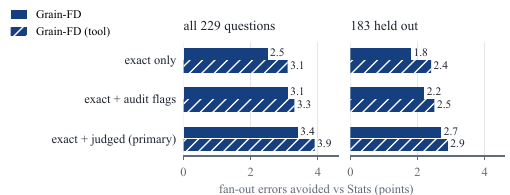}
\caption{BIRD-FanOut train: fan-out errors avoided against \textsc{Stats} under three metrics, two of them without LLM judges. Left: all 229 questions; right: the 183 held-out questions. Hatched: tool delivery.}
\label{fig:robust}
\end{figure}

\begin{figure}[t]
\centering
\includegraphics[width=\columnwidth]{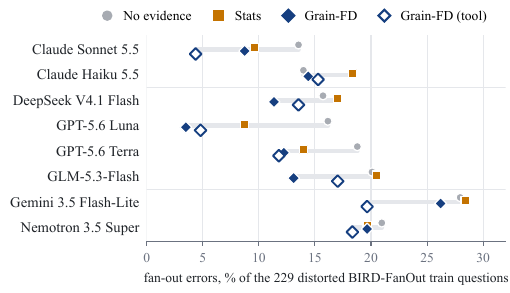}
\caption{Fan-out errors per model on BIRD-FanOut train (one run per question and arm). Tiers are separated by lines: low-, mid- and high-error. Against no evidence, the Grain-FD arms cut errors most for Luna, Sonnet and Terra; Haiku and Nemotron barely change.}
\label{fig:models}
\end{figure}

\textbf{Held-out questions, models and databases.} On the 183 held-out questions, the Grain-FD arms
avoid 2.7 and 2.9 points more than the statistics (both p $\le$ .001); on the
46 design questions, selected on \textsc{FD}'s outcomes, the effect is larger (6.0 and 8.2 points).
\textsc{Grain-FD} has fewer fan-out errors than the statistics for 7 of 8 models (one tie) and \textsc{Grain-FD (tool)}
for all 8, and resampling the 46 databases instead of the questions keeps both effects
(intervals {[}1.6, 5.3{]} and {[}1.9, 6.1{]} points). Against no evidence, fan-out errors fall most for Luna, Sonnet and Terra and barely
change for Nemotron and Haiku (Figure~\ref{fig:models}); Haiku's traces suggest it sees the repetition and
follows the hint's row-level formula (\S{}6.4). With one run per question, per-model differences are
descriptive.

\textbf{Judge-free metrics and judge reliability.} Two metrics that use no judge lead to the same conclusions (Figure~\ref{fig:robust}): exact fan-out errors alone, and exact ones plus other answers the audit flags as
definite fan-out. The judges add over-counts that the exact match misses, but they do not create the effect: among the
wrong answers that are not exact fan-out errors, the share the judges call fan-out falls with the
evidence (20.6\% without evidence, 14.4\% with \textsc{Grain-FD}) just as the share the audit flags does (11.8\%,
7.1\%).
Counting also the rare answers the judge calls grouped at the wrong grain (about 1\% of runs) as grain
errors gives the same result (3.1 and 3.7 points beyond the statistics, 2.4 and 2.7 on held-out
questions, p $\le$ .007).

The first judge, Gemini 3.6 Flash, sees the question and hint, the schema, both golds with their
results and the answer with its result, but not the audit's flag. Claude Opus 5.5 agreed with it on
98\% of a random sample (Cohen's $\kappa$ \cite{Cohen1960} 0.91) and settled every answer where Gemini and the
audit disagreed. Five Opus agents that could query each database, not shown the labels, matched them
on 47 of 50 answers ($\kappa$ 0.88), and their few disagreements suggest that judged fan-out, if anything,
slightly undercounts. These checks are not independent, because both judges saw the same prompt,
which says the question was flagged for over-counting, and Opus both settles disagreements and runs
the agents. The gold review (\S{}4) raised accuracy without evidence by 11 points but moved these
fan-out results by at most 0.5 points. Appendix D gives the full checks.

\textbf{Repetition counts or FDs? (exploratory)} The repetition counts alone give about half of the cut:
\textsc{Stats+Rep} avoids 2.0 points beyond the statistics (p = .013), 58\% of \textsc{Grain-FD}'s cut (95\%
interval 21--91\%). The rest of the grain-aware evidence avoids 1.4 more points, and 2.0 with the tool,
also on held-out questions (1.5 and 1.6 points, p = .028 and .036); plain FDs add nothing
beyond the repetition counts. The tool's increment is the firmer one, as it remains on exact fan-out
errors alone and in 7 of 8 models. No test in this exploratory family
survives Holm correction.

\textbf{Accuracy and dev.} Accuracy understates these effects, because most errors on these questions are
not fan-out and an avoided fan-out error often becomes another error (\S{}6.4). Against no evidence,
\textsc{Grain-FD} and \textsc{Grain-FD (tool)} raise accuracy from 60.3\% by 3.0 and 2.8 points (p = .003 and .009); on held-out questions
the gain is 1.8 points and not significant. Against the statistics they gain about 2 points (p = .031
and .07), mostly in the low- and mid-error tiers, and 0.6 points on held-out questions (not
significant), so we claim no accuracy gain of the FDs over the statistics. No arm raises the
high-error models' accuracy by more than 1.5 points, although \textsc{Grain-FD (tool)} cuts their fan-out errors by
5.5 points against no evidence. On dev,
only the 32 questions still distorted in the cleaned release can show fan-out, and they are too few
to test (fan-out errors: no evidence 38, \textsc{Grain-FD (tool)} 25 of 256 runs, p = .30); no accuracy
comparison on dev is significant. Only 13 of 85 dev answers that reproduce an old gold's result copy
its SQL nearly verbatim, so memorization of the public old gold plays a small role.

\subsection{Why does evidence help less on real data?}

\begin{figure}[t]
\centering
\includegraphics[width=\columnwidth]{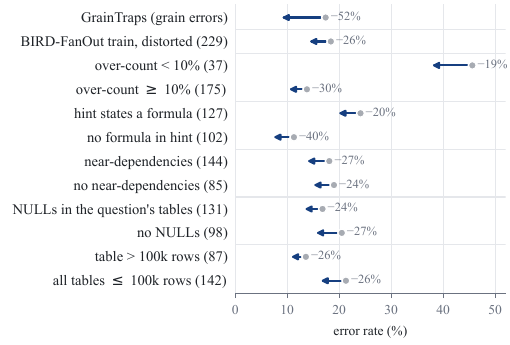}
\caption{Fan-out error rate without evidence (dot) and with \textsc{Grain-FD} (arrowhead), by question group (number of questions); top row: grain errors on GrainTraps. The arrow's length is the drop in percentage points; the label is the relative cut, the drop as a share of the rate without evidence (18.4\% $\rightarrow$ 13.6\% is $-$26\%). Over-count groups cover the 212 questions with a trap query, which is needed to measure the over-count. Over-count size and hint formulas matter; NULLs, near-dependencies and table size do not.}
\label{fig:splits}
\end{figure}

\textbf{\textsc{Grain-FD} halves grain errors on GrainTraps but cuts fan-out errors on BIRD by only a quarter,
mostly because of the questions and because some data barely repeats.} From no evidence to
\textsc{Grain-FD}, grain errors fall from 17.4\% to 8.3\% of runs on GrainTraps, where nearly every error is a grain
error, and fan-out errors fall from 18.4\% to 13.6\% on BIRD train. Figure~\ref{fig:splits} splits BIRD's 229
distorted questions by properties of the data and of the questions.

\emph{Repetition, size and noise.} When an over-count is small, the evidence gives little warning. For 17\%
of the BIRD questions with a trap query, the over-count changes the answer by less than 10\% (on
GrainTraps, for 4\%). In these questions few rows repeat, so the evidence reports close to one row per
value; their fan-out errors are frequent, and \textsc{Grain-FD} removes only 19\% of them, against 30\% for
larger over-counts. Other properties of the data matter little. BIRD's tables are larger and messier
than the trap tables (median 3,118 rows against 70, NULLs in a third of them, and in 30\% a dependency
that holds except for a few rows, which the trap tables never have), yet \textsc{Grain-FD} cuts fan-out errors
by 24--27\% with and without each of these.

\emph{The questions.} BIRD's questions differ from GrainTraps in two ways that matter more.

\begin{enumerate}
\def\labelenumi{\arabic{enumi}.}
\item
  \textbf{Hints prescribe the distorted formula.} Most BIRD questions come with a hint (BIRD's ``evidence'' field), written by its annotators for the
  original gold. For 127 of the 229 distorted train
  questions the hint states that gold's formula, such as ``average price per car = avg(price) where
  country = `Japan'\,''. With such a hint, fan-out errors are twice as frequent (24.1\% against 11.3\% of
  runs), and \textsc{Grain-FD} cuts them by 20\% instead of 40\%. GrainTraps has no hints. A Claude agent read 40
  \textsc{Grain-FD} runs that still made an exact fan-out error, five per model. In 21 the model saw the
  repetition and still chose the row-level reading, 15 of them with a hint formula (``The hint
  explicitly says avg(price) \ldots{} I'll go with the join answer'', GLM).
\item
  \textbf{A few hard, sometimes debatable questions carry many errors.} 22 of the 229 questions account
  for 45\% of the fan-out errors without evidence and 52\% with \textsc{Grain-FD}, while on GrainTraps no
  question keeps a grain error in most of its \textsc{Grain-FD} runs. In 11 of these 40 runs, the
  question or its corrected gold allows another reading.
\end{enumerate}

\emph{Accuracy.} A third difference explains why accuracy rises less, but not why fewer fan-out errors are
avoided. On BIRD 21\% of runs fail for other reasons (GrainTraps: 3\%), so an avoided error often becomes
another error. Of the runs that fall into the trap without evidence (an exact fan-out error on BIRD, a grain error
on GrainTraps), \textsc{Grain-FD} turns 58\% into correct answers on GrainTraps but
only 36\% on BIRD.

\emph{The evidence itself.} In 13 of the 40 runs, the evidence fell short. The needed fact was
missing (one column labelled a dimension instead of an identifier lost its repetition counts), buried
in over 100 lines of evidence, or stated and not used.

\subsection{When repeated rows should be counted}

Evidence that says rows repeat could push a model to deduplicate rows a question means to count. We
test this where counting the repeated rows is right: on GrainTraps' three control questions, which
every model solves in every run and arm, and on the 191 intended questions of BIRD-FanOut (dev and
train). With evidence in the prompt, no model is significantly less accurate on the intended questions than
without evidence (the largest drop is 1.6 points). With the tool arms, two models do
deduplicate where they should not. Against the statistics arm, Nemotron 3.5 Super loses 10.5 points
with \textsc{Grain-FD (tool)} and GPT-5.6 Luna 5.3, and the other models show no consistent loss. Against no evidence these losses are smaller and not significant, and they come from per-model
comparisons, not from the six headline tests (\S{}6), so they signal a risk of tool delivery rather than
an established harm. With \textsc{FD (tool)},
Nemotron loses 7.9 points and Luna 5.3. For the high-error tier, the losses against the statistics are
4.7 points (p = .037) and 4.2 (p = .058), not significant after correction; against no evidence they
are $-$2.1 and $-$1.6 points, and before the gold review the first was $-$4.2, so part of it was gold
error.

\subsection{The remaining errors}

\begin{figure}[t]
\centering
\includegraphics[width=\columnwidth]{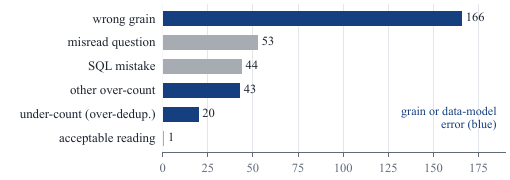}
\caption{Causes of the 327 distinct ``other'' trap answers (Opus judge; full-schema runs). Blue: causes the Opus judge counts as grain or data-model errors (229).}
\label{fig:causes}
\end{figure}

In the full-schema runs, most GrainTraps answers scored ``other'' (\S{}5.1) were still grain errors or
other misreadings of how the tables relate, which we call \emph{data-model errors} (Figure~\ref{fig:causes}). Two judges,
Claude Opus 5.5 (also a subject model in those runs) and GPT-5.6 Terra, classified each distinct
``other'' answer from 359 runs, seeing the question, its trap explanation and solution, the schema and
both results. They agree on the cause for 81\% of the answers ($\kappa$ = 0.72) and on whether it is a grain
or data-model error for 86\%.

Both judges call 210 of the 327 a grain or data-model error, so counting grain errors alone
understates how often models misread the data model. All but 8 wrong-grain answers are
rank-then-attach questions, where vague wording leaves some room. Four of the five ``other''
answers that Opus 5.5 gave as a subject model apply DISTINCT to line rows without the line key and so
merge lines that are really distinct, the mirror image of fan-out. Traces of open models given
evidence show three reasons it goes unused: per-table facts are not combined across a join; a tool
call scoped to a few columns leaves out the one that carries the measure's grain; and a prior belief
overrides the evidence (``assume each product has a single division'' after the tool
said otherwise).

\subsection{The audit as an offline verifier}

The audit can also check a model's answer before anyone uses it. In the full-schema runs we ran it on
every distinct submitted answer (2,161 on BIRD-FanOut dev, 1,807 on 150 ordinary Mini-Dev questions).
It catches most over-counts, flagging 70--75\% of the answers that reproduce a distorted gold (a
definite flag on 12--26\%, an ambiguous or repeated-values flag on the rest). It rarely flags correct
answers: 1\% on the dev questions with fan-out and 5\% on the ordinary questions, where only 0.3\% get a
definite flag. It cannot judge intent, though. On intended questions it flags 53\% of correct answers,
because a flag says that rows repeat, not whether the question meant them once, which the model or
user must decide.

\subsection{Cost}

The tool arms add about two tool calls per run. The evidence arms read up to twice the input tokens
of no evidence. The median per BIRD train run is 4.4k tokens without evidence, 4.9k with \textsc{Stats}, 5.4k
with \textsc{Stats+Rep}, 7.8k with \textsc{FD} and \textsc{FD (tool)}, 8.7k with \textsc{Grain-FD} and 7.7k with
\textsc{Grain-FD (tool)}. \textsc{Grain-FD} is the longest because it adds repetition counts, NULL-aware FDs and
measure grains, over 100 lines for some BIRD tables. All evidence is computed once per table and
cached; on BIRD train's largest databases (up to 5 GB), the first \textsc{Grain-FD} profile of a table takes
up to 27 minutes.

\section{Discussion}

\textbf{Lesson 1: Gold queries need a data-grounded review of the whole query.} Where BIRD's gold is distorted, a model that avoids the trap is scored wrong. The review must also
cover the hints, which BIRD's annotators wrote for the distorted gold and which often prescribe the
over-count (127 of 229 distorted train questions carry a formula hint); a model that sees the
repetition and follows such a hint resolves a conflict between two instructions rather than ignoring
the evidence. Fixing only the
fan-out left another BIRD error in one in five of our corrected train golds, and fixing these raised
measured accuracy on the same questions by 11 points (\S{}6.3). The review itself needs the data,
because two reviewers can agree on a fix that returns nothing.

\textbf{Lesson 2: Evidence should state the grain.} Per-column counts already reveal keys,
repetition (4,185 lines for 1,200 orders in Figure~\ref{fig:ladder}) and gaps. Correct FD labels (``can have many'') add
nothing beyond them on GrainTraps, or, in our exploratory control, beyond the repetition counts on
BIRD. What helps is evidence that
supplies the inputs of the grain condition (\S{}2.2): the grain of each measure and how many rows a join
repeats, together with where a blank breaks a dependency. MCI-SQL's small effect for dependency labels
\cite{MCI-SQL}, measured on accuracy, matches our accuracy results; measuring fan-out errors directly shows
a clearer effect.

\textbf{Lesson 3: Fan-out errors should be measured directly.} On real questions most errors are not
fan-out, and an over-count may also differ in other ways, so accuracy and exact matches understate
what evidence does. Judging each wrong answer for fan-out is feasible, since LLM judges, the audit and data-probing
agents agree closely (\S{}6.3), though not independently.

\textbf{Future work: facts about the model's own query.} Per-table facts leave the model to combine them
across joins, which it often fails to do, while checking the query against the data catches most
exact fan-out errors (\S{}6.7). A next step reports, for each relation the model's query uses, how its joins
repeat rows and the grain of each aggregated measure, inferred from per-table dependencies by
rules and computed from the data only where the rules leave it open. Unlike ontologies and semantic
layers, which need someone to declare the model, such facts come from the data.

\textbf{Limitations.}

\begin{itemize}
\item
  \emph{Scope.} Beyond BIRD, we audited only benchmarks that run on SQLite (Spider 2.0's BigQuery and
  Snowflake databases were not audited), and BEAVER's larger database only through a random sample of
  200 of its 1,186 definite flags. The audit checks a join to a subquery only statically, so EHRSQL
  \cite{EHRSQL} (its MIMIC-IV version), where every join that could repeat rows goes to a subquery, could
  not be tested. GrainTraps is small and synthetic, generated by a Claude agent,
  kept only if strong models could solve its questions, and has no held-out split. Without evidence
  the two Claude models make the fewest grain errors on it (3 and 5 of 168 runs, against 18 to 70 for
  the others), so comparisons across models may favour them; the arm effects are measured within each
  model. We use one dialect (SQLite; the audit uses \texttt{rowid}), two to four models per tier and one run
  per BIRD question (on GrainTraps, 14.5\% of three-run cells in the six main arms have mixed outcomes),
  and show the model only the tables its reference queries read, which makes the task easier than
  in practice.
\item
  \emph{Design.} The Grain-FD rules were developed on the design set (GrainTraps and 46 BIRD train
  questions selected on \textsc{FD}'s outcomes). They are generic, with no question-specific content, and the
  183 held-out questions test them, but their trap results are in-sample, no new trap questions were
  written afterwards, and their four elements were not ablated one by one. The measure grain is a
  single column, so a measure repeated per pair of columns (6 of 650 BIRD measures checked) is
  reported as varying by row. The arms ran on different days in November 2026 with the same code, \textsc{Stats+Rep} last
  with no same-day repeat of the others, so provider-side model updates could shift comparisons
  between arms, in particular the split between repetition counts and FDs. \textsc{Stats+Rep} was not run on dev, and its text is not a part of \textsc{Grain-FD}'s, so their difference is
  not a clean ablation. \textsc{FD} and \textsc{FD (tool)} differ in both wording and delivery, and the arms differ
  in length (\S{}6.8).
\item
  \emph{Labels.} No person labelled or checked any data. Column roles and BIRD labels come from LLM judges
  checked by Claude agents only on samples (one role error removed a needed fact). BIRD's fan-out
  errors rest on LLM judges that agree closely with each other, the audit and data-probing agents
  (the same model as the second judge), and the metric was chosen after the runs, once the gold review
  showed how many over-counts an exact match misses; the metrics that use no judge agree (Figure~\ref{fig:robust}). The gold review also rests on LLM judges, two of them also subject models (Claude
  Haiku 5.5 in the first stage, GPT-5.6 Terra in the second). A Claude agent checked it on the 21
  golds that prompted it and 5 accepted fixes (4 clearly right, 1 defensible), and every fix that
  returned no answer was checked on the data. The gold review left 40 questions out of BIRD-FanOut, and 17 distorted questions have no trap query, so
  their fan-out errors can be judged but never matched exactly. The runs were shown the tables the
  original golds read; no fixed gold reads any other table. The trace analysis of \S{}6.4 is qualitative,
  one Claude agent's coding pass over 40 runs. On the other benchmarks of \S{}4, both labels of a flag
  come from Claude agents, so no second model family checks them; the agents disagree on 29 of 484
  flags, 12 of them on one Advising gold whose correction changes only the row order.
\end{itemize}

\section{Conclusion}

Fan-out errors are silent and common in text-to-SQL, and one condition governs them: a
duplicate-sensitive aggregate is correct when each entity at its measure's grain appears once in the
relation it reads. The condition's inputs, the measure's grain and the multiplicity that joins induce,
are properties of the data that a schema does not show. Tested on the data, the condition exposes distorted
gold in a few percent of BIRD's questions. Supplied to LLM agents as grain-aware FDs, its inputs cut
LLM-judged fan-out errors on real questions from 18.4\% of runs without evidence, and 17.0\% with
statistics, to 13.1--13.6\%; on held-out questions, from 16.1\%
and 15.1\% to 12.2--12.4\%. Accuracy rises by only about 3 points, because the main obstacles that remain are
other errors, hints that prescribe the distorted formula, and over-counts too small to show in the
data. Checking the model's own query against the data, as the audit does, is the natural next
step. BIRD-FanOut, GrainTraps, the audit and the evidence code are available at \url{https://github.com/ahmed-mahran/misqal}.

\bibliographystyle{ACM-Reference-Format}
\bibliography{refs}

\appendix
\section{A second example of the evidence}

\begin{figure*}[t]
\centering
\includegraphics[width=\textwidth]{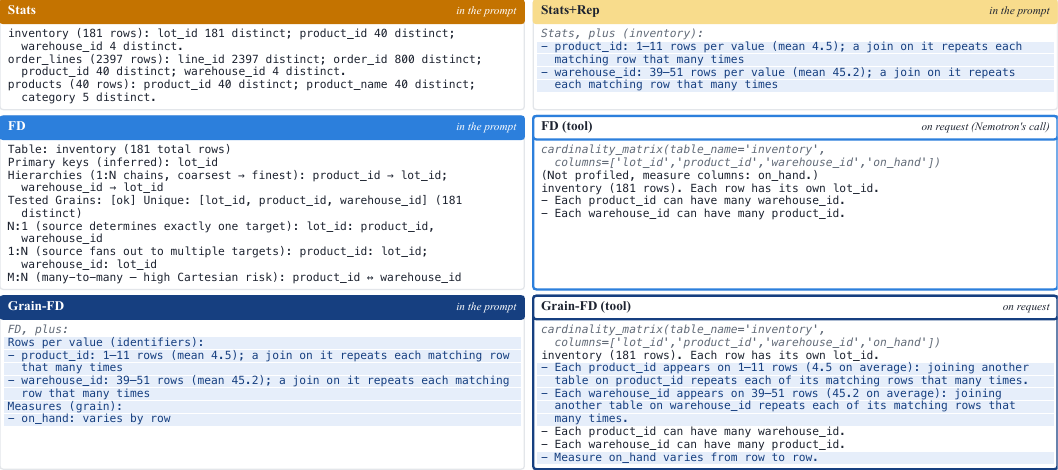}
\caption{The six evidence arms for a second GrainTraps question (``Which products have been ordered in a larger total quantity than the stock we hold of them?'', table \texttt{inventory}), copied from the run traces of GPT-5.6 Terra (the \textsc{FD (tool)} call is Nemotron 3.5 Super's), with markdown marks dropped and the many-to-many arrow drawn as $\leftrightarrow$. Rows: statistics, FDs and grain-aware FDs; columns: in the prompt and by tool. Blue: what an arm adds over its reference arm; grey italics: the model's tool call.}
\label{fig:ladder2}
\end{figure*}

Figure~\ref{fig:ladder} shows the evidence for a \texttt{shop} question. Figure~\ref{fig:ladder2} shows it for a
question from \texttt{supply}: ``Which products have been ordered in a larger total quantity than the stock
we hold of them?'' The known tables are \texttt{products}, \texttt{order\_lines} and \texttt{inventory}; the measures
\texttt{unit\_price}, \texttt{quantity} and \texttt{on\_hand} are left out of the statistics and the dependency profiles.

\section{Prompt (shared by all arms)}

Every run receives the system prompt and user message below; line breaks are reflowed. Grey text is
a placeholder (Figure~\ref{fig:ladder} shows the evidence blocks); the blue instruction appears only in
the two tool arms.

\begin{promptbox}{System prompt}
\begin{Verbatim}[fontsize=\scriptsize,breaklines,breaksymbolleft={},breakindent=0pt,commandchars=\\\{\}]
You answer questions about a SQLite database using the `execute_sql` tool. Every number you report must come from a query you ran in this turn. If a query fails, check the column and table names against the schema below and fix them. Do not assume how tables and columns relate from their names or from what seems typical: check it in the data, and treat a relationship you have not checked as unknown. Finish by calling `submit_answer` with the exact SQL whose result answers the question and a short answer in words. The submitted SQL must select exactly what the question asks for -- no extra columns (e.g. no counts or ids unless asked) -- with one value per cell (no lists or concatenated values in a column).
\end{Verbatim}
\begin{promptadd}
\begin{Verbatim}[fontsize=\scriptsize,breaklines,breaksymbolleft={},breakindent=0pt,formatcom=\color{promptarm}]
## Cardinality evidence (MANDATORY)
Before writing your final SQL you MUST run `cardinality_matrix` on EVERY table your query reads, passing the columns your query relies on. Base every relationship between columns that your query depends on, on its result. A submission that reads a table you have not run `cardinality_matrix` on is rejected.
\end{Verbatim}
\end{promptadd}
\begin{Verbatim}[fontsize=\scriptsize,breaklines,breaksymbolleft={},breakindent=0pt,commandchars=\\\{\}]
## Database schema
\textcolor{promptmuted}{<CREATE TABLE statements of the shown tables>}
\end{Verbatim}
\end{promptbox}
\begin{promptbox}{User message}
\begin{Verbatim}[fontsize=\scriptsize,breaklines,breaksymbolleft={},breakindent=0pt,formatcom=\color{promptmuted}]
<question>

<evidence block of the Stats, Stats+Rep, FD and Grain-FD arms>
\end{Verbatim}
\end{promptbox}

\section{Per-model results}

GrainTraps, 168 non-control runs per cell, correct / grain errors (Table~\ref{tab:2}).

\par\medskip\noindent\begin{minipage}{\columnwidth}\centering\scriptsize\setlength{\tabcolsep}{1.5pt}
\captionof{table}{Per model, GrainTraps (168 non-control runs per cell): correct / grain errors. Bold: fewest grain errors in the row (ties all bold).}\label{tab:2}
\begin{tabular}{@{}lrrrrrrr@{}}
\toprule
\textbf{model} & \textbf{\shortstack[l]{no\\evidence}} & \textbf{\textsc{Stats}} & \textbf{\textsc{Stats+Rep}} & \textbf{\textsc{FD}} & \textbf{\shortstack[l]{\textsc{FD}\\\textsc{(tool)}}} & \textbf{\textsc{Grain-FD}} & \textbf{\shortstack[l]{\textsc{Grain-FD}\\\textsc{(tool)}}} \\
\midrule
Claude Sonnet 5.5 & 165 / 3 & \textbf{168 / 0} & \textbf{168 / 0} & 164 / 4 & \textbf{168 / 0} & \textbf{168 / 0} & \textbf{168 / 0} \\
Claude Haiku 5.5 & 162 / 5 & 165 / 2 & 165 / 1 & 158 / 4 & 165 / 2 & \textbf{166 / 0} & \textbf{165 / 0} \\
DeepSeek V4.1 Flash & 147 / 18 & 151 / 15 & 153 / 11 & 159 / 8 & 154 / 11 & \textbf{166 / 0} & 165 / 1 \\
GPT-5.6 Luna & 150 / 18 & 156 / 12 & 152 / 14 & 152 / 16 & 148 / 18 & 161 / 5 & \textbf{163 / 3} \\
GPT-5.6 Terra & 145 / 21 & 160 / 7 & 158 / 9 & 153 / 13 & 148 / 19 & \textbf{167 / 0} & 162 / 5 \\
GLM-5.3-Flash & 130 / 32 & 142 / 18 & 147 / 12 & 140 / 13 & 129 / 24 & \textbf{157 / 7} & 142 / 15 \\
Gemini 3.5 Flash-Lite & 88 / 67 & 101 / 52 & 114 / 35 & 112 / 44 & 110 / 42 & 123 / 39 & \textbf{126 / 31} \\
Nemotron 3.5 Super & 83 / 70 & 88 / 66 & 89 / 63 & 90 / 61 & 95 / 60 & 95 / 61 & \textbf{106 / 49} \\
\bottomrule
\end{tabular}
\end{minipage}\par\medskip

BIRD-FanOut train, 229 distorted questions (reviewed golds), correct / exact fan-out errors (the answer
reproduces the distorted gold or the agreed trap query) (Table~\ref{tab:3}).

\par\medskip\noindent\begin{minipage}{\columnwidth}\centering\scriptsize\setlength{\tabcolsep}{1.5pt}
\captionof{table}{Per model, BIRD-FanOut train (229 distorted questions): correct / exact fan-out errors. Bold: fewest exact fan-out errors in the row.}\label{tab:3}
\begin{tabular}{@{}lrrrrrrr@{}}
\toprule
\textbf{model} & \textbf{\shortstack[l]{no\\evidence}} & \textbf{\textsc{Stats}} & \textbf{\textsc{Stats+Rep}} & \textbf{\textsc{FD}} & \textbf{\shortstack[l]{\textsc{FD}\\\textsc{(tool)}}} & \textbf{\textsc{Grain-FD}} & \textbf{\shortstack[l]{\textsc{Grain-FD}\\\textsc{(tool)}}} \\
\midrule
Claude Sonnet 5.5 & 151 / 26 & 162 / 18 & 168 / 11 & 159 / 20 & \textbf{172 / 8} & 162 / 15 & \textbf{171 / 8} \\
Claude Haiku 5.5 & \textbf{152 / 22} & 138 / 31 & 134 / 32 & 145 / 29 & 145 / 29 & 149 / 29 & 150 / 28 \\
DeepSeek V4.1 Flash & 140 / 27 & 138 / 28 & 147 / 25 & 144 / 23 & 149 / 21 & 148 / 20 & \textbf{150 / 19} \\
GPT-5.6 Luna & 148 / 22 & 158 / 15 & 168 / 8 & 162 / 9 & 160 / 13 & \textbf{171 / 5} & 164 / 7 \\
GPT-5.6 Terra & 142 / 28 & 153 / 22 & 144 / 25 & \textbf{153 / 20} & 144 / 24 & \textbf{150 / 20} & \textbf{147 / 20} \\
GLM-5.3-Flash & 134 / 37 & 133 / 33 & 137 / 21 & \textbf{142 / 19} & 129 / 30 & 145 / 21 & 130 / 24 \\
Gemini 3.5 Flash-Lite & 122 / 45 & 120 / 48 & 127 / 37 & 125 / 46 & 122 / 43 & 121 / 42 & \textbf{133 / 34} \\
Nemotron 3.5 Super & 116 / 29 & 120 / 30 & 116 / 32 & 105 / 36 & 114 / 30 & \textbf{114 / 27} & 112 / 29 \\
\bottomrule
\end{tabular}
\end{minipage}\par\medskip

BIRD-FanOut train, fan-out errors (exact plus other answers judged to over-count; \S{}6.3) (Table~\ref{tab:4}).

\par\medskip\noindent\begin{minipage}{\columnwidth}\centering\scriptsize\setlength{\tabcolsep}{1.5pt}
\captionof{table}{Per model, BIRD-FanOut train, fan-out errors. Bold: fewest in the row.}\label{tab:4}
\begin{tabular}{@{}lrrrrrrr@{}}
\toprule
\textbf{model} & \textbf{\shortstack[l]{no\\evidence}} & \textbf{\textsc{Stats}} & \textbf{\textsc{Stats+Rep}} & \textbf{\textsc{FD}} & \textbf{\shortstack[l]{\textsc{FD}\\\textsc{(tool)}}} & \textbf{\textsc{Grain-FD}} & \textbf{\shortstack[l]{\textsc{Grain-FD}\\\textsc{(tool)}}} \\
\midrule
Claude Sonnet 5.5 & 31 & 22 & 15 & 24 & \textbf{10} & 20 & \textbf{10} \\
Claude Haiku 5.5 & \textbf{32} & 42 & 42 & 39 & 37 & 33 & 35 \\
DeepSeek V4.1 Flash & 36 & 39 & 35 & 34 & 28 & \textbf{26} & 31 \\
GPT-5.6 Luna & 37 & 20 & 12 & 15 & 18 & \textbf{8} & 11 \\
GPT-5.6 Terra & 43 & 32 & 36 & 30 & 31 & 28 & \textbf{27} \\
GLM-5.3-Flash & 46 & 47 & 36 & \textbf{26} & 37 & 30 & 39 \\
Gemini 3.5 Flash-Lite & 64 & 65 & 53 & 60 & 60 & 60 & \textbf{45} \\
Nemotron 3.5 Super & 48 & 45 & 47 & 52 & 49 & 45 & \textbf{42} \\
\bottomrule
\end{tabular}
\end{minipage}\par\medskip

\section{Supporting runs and checks}

\begin{itemize}
\item
  \textbf{Full-schema runs} (\S{}5.2). The eight models of \S{}5.2 ran no evidence, \textsc{FD} and
  \textsc{FD (tool)} with the whole database shown and measures included in the evidence. On
  GrainTraps, evidence cut grain errors by 5--12 points for the four smaller models and by 4--5 points
  for the four larger ones, almost all on NULL splits. On the 150 ordinary Mini-Dev questions it did no
  harm (accuracy with \textsc{FD (tool)} +1.0 points {[}$-$2.0, +4.2{]}).
\item
  \textbf{Design set} (\S{}5.2). The Grain-FD rules were developed on a pilot of 350 runs: the GrainTraps
  questions where \textsc{FD} did worse than \textsc{Stats}, the NULL-split questions, and 49
  BIRD train questions where \textsc{FD} changed Haiku's or Luna's outcome (46 remain after the
  gold review). The pilot runs are part of the reported results; all other Grain-FD runs came after
  the rules were fixed.
\item
  \textbf{Gold review} (\S{}4). Of the 421 train golds, 80 were fixed (56 distorted, 24 intended) and 37 were
  left out: 27 disputed by the judges, 10 broken or without a fan-out signal. Twelve accepted fixes
  returned nothing, 0 or NULL and were checked on the data: two were corrected, nine were left out (an
  answer of 0 or NULL cannot show fan-out, or the question was ambiguous) and one was kept.
\item
  \textbf{Robustness to the gold review} (\S{}6.3). The review came after all runs but those of \textsc{Stats+Rep}, which were re-scored. It
  barely moved the fan-out results: \textsc{Grain-FD} cuts fan-out errors by 3.9 points beyond
  \textsc{Stats} on the original golds and by 3.4 after the review, \textsc{Grain-FD (tool)}
  by 4.1 and 3.9. It moved accuracy more, because gold errors had marked correct answers wrong: without
  evidence, from 49.1\% to 60.3\% on distorted questions and from 68.7\% to 78.8\% on intended ones. On the
  original golds, the accuracy gains over \textsc{Stats} were +1.9 points (p = .018) and +1.4
  (p = .20). After the review, the judges still call 6.5\% of runs acceptable answers that the gold does
  not match; we count these as wrong.
\item
  \textbf{Judged fan-out} (\S{}6.3). Where Gemini and the audit disagreed, Opus mostly sided with Gemini: it
  confirmed 208 of the 239 answers Gemini called fan-out that the audit did not flag, and rejected 16
  of the 20 flagged answers Gemini did not call fan-out. Given answers equal to the distorted gold,
  Gemini called 294 of 300 fan-out. Re-judging every distinct answer of all seven arms in one batch
  agreed with the original labels on 98.7\% and changed no conclusion.
\item
  \textbf{Other benchmarks} (\S{}4). Every gold query was audited on its own database; Spider's databases that
  ship without rows are left out. ATIS, Advising and BEAVER were moved to SQLite from their MySQL
  releases and FIBEN from Db2, keeping only golds that run; four FIBEN golds needed a GROUP BY on an
  outer column dropped, and BEAVER's variances and standard deviations were rebuilt from sums and
  counts, which keeps their sensitivity to repeated rows. Two Claude agents labelled each definite flag without seeing each
  other's label; a gold is distorted if both do, and a change in row order alone does not count. For
  BEAVER's template-generated warehouse questions, a random 200 of the 1,186 definite flags were
  labelled and the count is scaled, with a 95\% Wilson interval. Spider 2.0-Lite counts only its 24
  SQLite questions with public gold (Table~\ref{tab:5}).
\end{itemize}

\par\medskip\noindent\begin{minipage}{\columnwidth}\centering\scriptsize\setlength{\tabcolsep}{1.5pt}
\captionof{table}{Fan-out in the gold of other benchmarks (\S4): golds, the share that join, definite flags, and golds distorted by the labels of two Claude agents (BIRD dev: two LLM judges), as a count and as a \% of all golds and of golds that join. BEAVER warehouse, generated: scaled from 200 labelled flags (95\% interval).}\label{tab:5}
\begin{tabular}{@{}lrrrlrr@{}}
\toprule
\textbf{Benchmark} & \textbf{Golds} & \textbf{Join} & \textbf{Definite} & \textbf{Distorted} & \textbf{\% golds} & \textbf{\% joined} \\
\midrule
BIRD dev (June 2024) & 1,534 & 74\% & 82 & 64 & 4.2 & 5.6 \\
Spider 1.0 dev & 1,034 & 39\% & 1 & 0 & 0 & 0 \\
Spider 1.0 test & 2,147 & 40\% & 20 & 4 & 0.19 & 0.46 \\
Spider 1.0 train & 6,900 & 39\% & 47 & 12 & 0.17 & 0.45 \\
KaggleDBQA & 272 & 16\% & 3 & 1 & 0.37 & 2.3 \\
Spider 2.0-Lite (SQLite) & 24 & 92\% & 2 & 1 & 4.2 & 4.5 \\
ATIS & 4,851 & 93\% & 0 & 0 & 0 & 0 \\
Advising & 4,334 & 85\% & 26 & 0 & 0 & 0 \\
FIBEN & 227 & 100\% & 1 & 0 & 0 & 0 \\
BEAVER warehouse, query logs & 92 & 100\% & 26 & 22 & 23.9 & 23.9 \\
BEAVER warehouse, generated & 5,266 & 100\% & 1,186 & 480 (403--562) & 9.1 & 9.1 \\
BEAVER network, generated & 842 & 100\% & 158 & 87 & 10.3 & 10.3 \\
\bottomrule
\end{tabular}
\end{minipage}\par\medskip
\onecolumn
\section{Full results tables}

These tables give the exact numbers behind Figures~\ref{fig:check}, \ref{fig:effects}, \ref{fig:robust} and \ref{fig:causes}. Comparisons read ``tested arm vs reference''; grain and fan-out errors are given as points avoided (positive is better), accuracy as points gained (Table~\ref{tab:6}).

\begin{table*}[!htbp]
\centering\footnotesize
\caption{GrainTraps without evidence: correct runs and grain errors out of 168 non-control runs per model, and the share of runs that ran a checking query (Figure~\ref{fig:check}). Bold: the best in each column.}
\label{tab:6}
\begin{tabular}{@{}llrrr@{}}
\toprule
\textbf{tier} & \textbf{model} & \textbf{correct} & \textbf{grain errors} & \textbf{runs that ran a checking query} \\
\midrule
low-error & Claude Sonnet 5.5 & \textbf{165} & \textbf{3} & \textbf{98\%} \\
low-error & Claude Haiku 5.5 & 162 & 5 & 95\% \\
mid-error & DeepSeek V4.1 Flash & 147 & 18 & 67\% \\
mid-error & GPT-5.6 Luna & 150 & 18 & 32\% \\
mid-error & GPT-5.6 Terra & 145 & 21 & 27\% \\
mid-error & GLM-5.3-Flash & 130 & 32 & 37\% \\
high-error & Gemini 3.5 Flash-Lite & 88 & 67 & 21\% \\
high-error & Nemotron 3.5 Super & 83 & 70 & 11\% \\
\bottomrule
\end{tabular}
\end{table*}

\begin{table*}[!htbp]
\centering\footnotesize
\caption{GrainTraps, grain errors avoided by the tested arm (points, 95\% interval), then base questions with fewer / more grain errors (a question and its vague twin count as one); for all tiers, permutation $p$ (Figure~\ref{fig:effects}, left).}
\label{tab:7}
\begin{tabular}{@{}lllll@{}}
\toprule
\textbf{comparison} & \textbf{low-error} & \textbf{mid-error} & \textbf{high-error} & \textbf{all} \\
\midrule
\textsc{Stats} vs no evidence & +1.8 {[}+0.0, +4.5{]}, 2/0 & +5.5 {[}+0.9, +11.5{]}, 9/4 & +5.7 {[}$-$0.6, +12.7{]}, 13/9 & +4.6 {[}+1.3, +8.6{]}, 15/7, p = .015 \\
\textsc{FD} vs \textsc{Stats} & $-$1.8 {[}$-$5.6, +0.0{]}, 0/1 & +0.3 {[}$-$5.7, +5.8{]}, 8/2 & +3.9 {[}$-$0.6, +8.0{]}, 13/6 & +0.7 {[}$-$3.6, +4.0{]}, 15/7, p = .78 \\
\textsc{FD (tool)} vs \textsc{Stats} & 0.0 {[}$-$1.2, +1.5{]}, 1/2 & $-$3.0 {[}$-$8.2, +0.8{]}, 5/8 & +4.8 {[}+0.0, +9.8{]}, 13/6 & $-$0.3 {[}$-$3.2, +2.0{]}, 12/8, p = .88 \\
\textbf{\textsc{Grain-FD} vs \textsc{Stats}} & +0.6 {[}+0.0, +1.8{]}, 1/0 & +6.0 {[}+1.8, +11.8{]}, 10/1 & +5.4 {[}$-$0.6, +11.5{]}, 15/6 & \textbf{+4.5 {[}+1.3, +8.3{]}, 15/6, p = .012} \\
\textbf{\textsc{Grain-FD (tool)} vs \textsc{Stats}} & +0.6 {[}+0.0, +1.8{]}, 1/0 & +4.2 {[}$-$0.1, +10.3{]}, 7/5 & +11.3 {[}+4.5, +18.5{]}, 16/4 & \textbf{+5.1 {[}+1.9, +8.9{]}, 15/5, p = .003} \\
\textsc{Stats+Rep} vs \textsc{Stats} & +0.3 {[}+0.0, +0.9{]}, 1/0 & +0.9 {[}$-$1.7, +3.4{]}, 8/3 & +6.0 {[}+1.2, +10.7{]}, 14/5 & +2.0 {[}+0.6, +3.4{]}, 15/6, p = .019 \\
\textsc{Grain-FD} vs \textsc{Stats+Rep} & +0.3 {[}+0.0, +0.9{]}, 1/0 & +5.1 {[}+0.6, +11.1{]}, 8/2 & $-$0.6 {[}$-$7.2, +5.7{]}, 11/8 & +2.5 {[}$-$0.6, +6.1{]}, 14/7, p = .19 \\
\textsc{Grain-FD (tool)} vs \textsc{Stats+Rep} & +0.3 {[}+0.0, +0.9{]}, 1/0 & +3.3 {[}$-$1.2, +9.6{]}, 7/5 & +5.4 {[}$-$0.6, +12.0{]}, 11/9 & +3.1 {[}+0.1, +6.7{]}, 10/6, p = .089 \\
\bottomrule
\end{tabular}
\end{table*}

\begin{table*}[!htbp]
\centering\footnotesize
\caption{GrainTraps, accuracy gain of the tested arm (points, 95\% interval), then base questions with more / fewer correct runs; for all tiers, permutation $p$.}
\label{tab:8}
\begin{tabular}{@{}lllll@{}}
\toprule
\textbf{comparison} & \textbf{low-error} & \textbf{mid-error} & \textbf{high-error} & \textbf{all} \\
\midrule
\textsc{Stats} vs no evidence & +1.8 {[}+0.0, +4.5{]}, 2/0 & +5.5 {[}+1.2, +11.0{]}, 12/5 & +5.4 {[}$-$0.9, +12.6{]}, 13/7 & +4.5 {[}+1.5, +8.3{]}, 16/5, p = .011 \\
\textsc{FD} vs \textsc{Stats} & $-$3.3 {[}$-$7.6, $-$0.6{]}, 0/5 & $-$0.7 {[}$-$6.7, +4.8{]}, 9/7 & +3.9 {[}$-$0.6, +8.2{]}, 13/8 & $-$0.2 {[}$-$4.4, +3.2{]}, 13/12, p = .94 \\
\textsc{FD (tool)} vs \textsc{Stats} & 0.0 {[}$-$1.2, +1.5{]}, 1/2 & $-$4.5 {[}$-$9.5, $-$0.4{]}, 5/11 & +4.8 {[}$-$0.6, +9.9{]}, 16/6 & $-$1.0 {[}$-$4.3, +1.5{]}, 13/10, p = .58 \\
\textbf{\textsc{Grain-FD} vs \textsc{Stats}} & +0.3 {[}$-$1.2, +1.9{]}, 2/2 & +6.2 {[}+1.8, +12.3{]}, 11/4 & +8.6 {[}+2.3, +15.2{]}, 15/6 & \textbf{+5.4 {[}+2.0, +9.2{]}, 16/7, p = .006} \\
\textbf{\textsc{Grain-FD (tool)} vs \textsc{Stats}} & 0.0 {[}$-$1.8, +1.8{]}, 1/1 & +3.4 {[}$-$1.3, +10.0{]}, 9/8 & +12.8 {[}+5.7, +20.0{]}, 18/3 & \textbf{+4.9 {[}+1.4, +8.9{]}, 16/4, p = .010} \\
\textsc{Stats+Rep} vs \textsc{Stats} & 0.0 {[}$-$0.9, +0.9{]}, 1/1 & +0.1 {[}$-$2.9, +3.2{]}, 8/9 & +4.2 {[}$-$0.6, +8.8{]}, 14/9 & +1.1 {[}$-$0.9, +3.1{]}, 15/10, p = .34 \\
\textsc{Grain-FD} vs \textsc{Stats+Rep} & +0.3 {[}$-$1.2, +2.0{]}, 2/2 & +6.1 {[}+1.1, +12.4{]}, 12/4 & +4.5 {[}$-$2.9, +10.9{]}, 16/7 & +4.2 {[}+0.7, +8.3{]}, 17/7, p = .045 \\
\textsc{Grain-FD (tool)} vs \textsc{Stats+Rep} & 0.0 {[}$-$1.5, +1.2{]}, 2/1 & +3.3 {[}$-$1.5, +9.8{]}, 10/9 & +8.6 {[}+1.8, +15.5{]}, 14/8 & +3.8 {[}+0.3, +8.0{]}, 13/7, p = .067 \\
\bottomrule
\end{tabular}
\end{table*}

\begin{table*}[!htbp]
\centering\footnotesize
\caption{BIRD-FanOut train (229 distorted questions, 1,832 runs per arm), fan-out errors avoided by the tested arm (points; positive is better), questions with fewer / more fan-out runs; for all tiers, 95\% interval and permutation $p$; tiers are descriptive (Figure~\ref{fig:effects}, right).}
\label{tab:9}
\begin{tabular}{@{}lrrrl@{}}
\toprule
\textbf{comparison} & \textbf{low-error} & \textbf{mid-error} & \textbf{high-error} & \textbf{all} \\
\midrule
\textsc{Stats} vs no evidence & $-$0.2, 8/12 & +2.6, 38/25 & +0.4, 23/20 & +1.4 {[}$-$0.1, +2.9{]}, 49/39, p = .087 \\
\textsc{FD} vs \textsc{Stats} & +0.2, 11/10 & +3.6, 39/13 & $-$0.4, 22/24 & +1.7 {[}+0.3, +3.2{]}, 48/30, p = .028 \\
\textsc{FD (tool)} vs \textsc{Stats} & +3.7, 22/6 & +2.6, 45/23 & +0.2, 25/24 & +2.3 {[}+0.8, +3.7{]}, 58/33, p = .004 \\
\textbf{\textsc{Grain-FD} vs \textsc{Stats}} & +2.4, 19/9 & +5.0, 43/12 & +1.1, 26/21 & \textbf{+3.4 {[}+1.9, +4.9{]}, 59/26, p \textless{} .001} \\
\textbf{\textsc{Grain-FD (tool)} vs \textsc{Stats}} & +4.1, 25/8 & +3.3, 40/19 & +5.0, 32/11 & \textbf{+3.9 {[}+2.2, +5.6{]}, 67/29, p \textless{} .001} \\
\textsc{Stats+Rep} vs \textsc{Stats} & +1.5, 13/7 & +2.1, 36/21 & +2.2, 26/20 & +2.0 {[}+0.5, +3.4{]}, 49/35, p = .013 \\
\textsc{Grain-FD} vs \textsc{Stats+Rep} & +0.9, 12/9 & +2.9, 34/14 & $-$1.1, 25/28 & +1.4 {[}+0.3, +2.6{]}, 54/28, p = .024 \\
\textsc{Grain-FD (tool)} vs \textsc{Stats+Rep} & +2.6, 17/6 & +1.2, 32/26 & +2.8, 35/20 & +2.0 {[}+0.6, +3.4{]}, 57/29, p = .007 \\
\bottomrule
\end{tabular}
\end{table*}

\begin{table*}[!htbp]
\centering\footnotesize
\caption{BIRD-FanOut train, fan-out errors avoided against the statistics with and without the LLM judges, and grain errors (points), questions with fewer / more fan-out runs, permutation $p$; all 229 distorted questions and the 183 held-out questions (Figure~\ref{fig:robust}).}
\label{tab:10}
\begin{tabularx}{\textwidth}{@{}Xllll@{}}
\toprule
\textbf{fan-out metric} & \textbf{\textsc{Grain-FD} vs \textsc{Stats}, 229} & \textbf{\textsc{Grain-FD (tool)} vs \textsc{Stats}, 229} & \textbf{\textsc{Grain-FD} vs \textsc{Stats}, 183} & \textbf{\textsc{Grain-FD (tool)} vs \textsc{Stats}, 183} \\
\midrule
exact only & +2.5, 44/18, p \textless{} .001 & +3.1, 46/19, p \textless{} .001 & +1.8, 31/14, p = .008 & +2.4, 32/15, p = .002 \\
exact + audit flags & +3.1, 57/25, p \textless{} .001 & +3.3, 59/27, p \textless{} .001 & +2.2, 42/21, p = .004 & +2.5, 43/23, p = .002 \\
exact + judged (primary) & +3.4, 59/26, p \textless{} .001 & +3.9, 67/29, p \textless{} .001 & +2.7, 44/21, p \textless{} .001 & +2.9, 49/27, p = .001 \\
grain errors (exact + judged fan-out or wrong grain) & +3.1, 60/31, p \textless{} .001 & +3.7, 66/34, p \textless{} .001 & +2.4, 45/25, p = .007 & +2.7, 48/30, p = .004 \\
\bottomrule
\end{tabularx}
\end{table*}

\begin{table*}[!htbp]
\centering\footnotesize
\caption{BIRD-FanOut train, fan-out errors per arm on the 183 held-out questions (1,464 runs per arm; Figure~\ref{fig:robust}). Bold: fewest.}
\label{tab:11}
\begin{tabular}{@{}lrrrrrr@{}}
\toprule
\textbf{questions} & \textbf{no evidence} & \textbf{\textsc{Stats}} & \textbf{\textsc{FD}} & \textbf{\textsc{FD (tool)}} & \textbf{\textsc{Grain-FD}} & \textbf{\textsc{Grain-FD (tool)}} \\
\midrule
183 held out & 236 & 221 & 204 & 189 & 181 & \textbf{179} \\
\bottomrule
\end{tabular}
\end{table*}

\begin{table*}[!htbp]
\centering\footnotesize
\caption{Causes of the ``other'' trap answers (Opus judge; full-schema runs; Figure~\ref{fig:causes}).}
\label{tab:12}
\begin{tabularx}{\textwidth}{@{}Xr@{}}
\toprule
\textbf{cause (Opus judge)} & \textbf{distinct answers} \\
\midrule
wrong grain (ranked or broken down at another grain than asked) & 166 \\
misread question & 53 \\
SQL mistake (20 are NULL handling) & 44 \\
over-count not matching the naive SQL & 43 \\
under-count: real rows removed as duplicates & 20 \\
acceptable reading the gold rejects & 1 \\
\bottomrule
\end{tabularx}
\end{table*}

\end{document}